\documentclass[prb,preprint,showpacs,preprintnumbers,amsmath,amssymb,longbibliography]{revtex4-1}

\usepackage{todonotes}
\usepackage{graphicx}
\usepackage{newtxtext}
\usepackage{newtxmath}
\usepackage{natbib}
\usepackage{hyperref}
\usepackage{mathtools}
\usepackage{commath}
\usepackage{tabularx}

\newcommand{\RomanNumeralCaps}[1]
\linenumbers

\renewcommand{\aa}{\mathbf{a}}
\newcommand{\bd}{\partial}
\newcommand{\dd}{\mathbf{d}}
\newcommand{\DD}{\mathcal{D}}
\newcommand{\DDD}{\boldsymbol{\mathcal{D}}}
\newcommand{\SSS}{\mathcal{S}}
\newcommand{\eeta}{\boldsymbol{\eta}}
\newcommand{\FF}{\mathbf{F}}

\newcommand{\nnu}{\boldsymbol{\nu}}

\newcommand{\ssigma}{\boldsymbol{\sigma}}
\newcommand{\rr}{\mathbf{r}}
\newcommand{\RR}{\mathbf{R}}
\renewcommand{\SS}{\mathbf{S}}
\newcommand{\ee}{\mathbf{e}}
\newcommand{\ff}{\mathbf{f}}
\newcommand{\xx}{\mathbf{x}}
\newcommand{\zz}{\mathbf{z}}

\newcommand{\uu}{\mathbf{u}}
\renewcommand{\vv}{\mathbf{v}}
\newcommand{\yy}{\mathbf{y}}
\renewcommand{\tt}{\mathbf{t}}
\newcommand{\pderiv}[2]{\frac{\partial #1}{\partial #2}}

\begin{document}


\title{Effects of Tunable Hydrophobicity on the Collective Hydrodynamics of Janus Particles under Flows}


\author{Szu-Pei Fu$^{1}$, Rolf Ryham$^{2}$, Bryan Quaife$^{3}$ and Y.-N. Young$^{4}$}
\affiliation{$^1$Department of Mathematics, Trinity College, Hartford, Connecticut 06106, USA\\
$^2$Department of Mathematics, Fordham University, Bronx, New York 10458, USA \\
$^3$Department of Scientific Computing, Florida State University, Tallahassee, Florida 32306, USA\\
$^4$Department of Mathematical Sciences, New Jersey Institute of Technology, Newark, New Jersey 07102, USA}

%


\date{\today}

\begin{abstract}
  Active colloidal systems with non-equilibrium self-organization is a
  long-standing, challenging area in biology. To understand how
  hydrodynamic flow may be used to actively control self-assembly of
  Janus particles (JPs), we use a model recently developed for the
  many-body hydrodynamics of amphiphilic JPs suspended in a viscous
  background flow (JFM, 941, 2022). We investigate how various
  morphologies arise from tuning the hydrophobic distribution of the
  JP-solvent interface. We find JPs assembled into uni-lamella,
  multi-lamella and striated structures. To introduce dynamics, we
  include a linear shear flow and a steady Taylor-Green mixing flow, and
  measure the collective dynamics of JP particles in terms of their (a)
  free energy from the hydrophobic interactions between the JPs, (b)
  order parameter for the ordering of JPs in terms of alignment of their
  directors, and (c) strain parameter that captures the deformation in
  the assembly. We characterize the effective material properties of the
  JP structures and find that the uni-lamellar structures increases
  orientation order under shear flow, the multilamellar structure
  behaves as a shear thinning fluid, and the striated structure
  possesses a yield stress. These numerical results provide insights
  into dynamic control of non-equilibrium active biological systems with
  similar self-organization.
\end{abstract}



\maketitle


\section{Introduction}
Janus particles are colloids combining two dissimilar chemical or
physical functionalities at their opposite
sides~\cite{KirillovaMarschelkeSynytska2019}. Self-propelling Janus
particles, for example, with a permanent biphasic asymmetry, have
emerged as a rich chemical platform for the exploration of active
matter~\cite{Meredithetal2022}.
In the absence of mobility and any
imposed flow, Janus particles (JPs) self-assemble
into oligomers of various geometries
and sizes
depending on the interactions between JPs and the 
viscous solvent~\cite{Bradley2017, KangHonciuc2018,
HongCacciutoLuijtenGranick2008},
with tunable functions for biomedical engineering
applications~\cite{GheisariSahfieeAbbasiEtAl2021_DMR,
LiuYangHuangEtAl2016_Angew, LiWangYaoEtAl2019_Nanoscale,
Bradley2016,Zarzaretal2015, KirillovaMarschelkeSynytska2019}.

When driven by an external flow, dynamic rearrangement of JPs emerges
naturally from the interactions between particles and fluid in colloidal
matter~\cite{RevModPhys.93.025008}. Such colloidal hydrodynamics
belong to a wide class of nonequilibrium self-organization in physics,
often with complexity and features similar to that of biological systems
such as living cells, bacterial baths, and animal
flocks~\cite{CollardGrosjeanVandewalle2020, Vutukuri2020}. A
long-standing challenge in fluid mechanics and material science is to
solve the ``inverse" problem of creating a model colloidal system that
will self-assemble into a prescribed
structure~\cite{PhysRevLett.128.256102}. For example, the framework of
geometrical frustration, which has been used to explain disordered
systems, could instead be used to design new, ordered systems, through
specific choices for the shapes or interactions of the
particles~\cite{Manoharan2015_Science}.

The surface of a Janus particle may have a dissimilar spatial
distribution of hydrophobicity~\cite{doi:10.1021/la503455h}. The
many-body hydrodynamics of amphiphilic Janus particles assembled as
vesicles suspended in a viscous fluid in the inertialess regime (zero
Reynolds number) has been studied using boundary integral numerical
simulations~\cite{Fu20,Fu2022_JFM}. The dynamics of the JP suspension
results from the combination of long-range hydrodynamic interaction and
non-local interactions between JPs through the distribution of a
hydrophobic attraction potential (HAP).
In a quiescent flow, numerical simulations of a JP suspension showed
self-assembly into micelles and bilayers of JPs that provide an
alternative means for computing the mechanical moduli of a colloidal
membrane~\cite{NaTr00, Fu20, KrFiGuKaHa13}. Under background flows, the
hydrodynamics of a JP vesicle (a self-enclosed bilayer of JPs) exhibit
many familiar behaviors of a vesicle: elongation and alignment along the
extensional direction, tank-treading, and rupture of a vesicle under
shear flow~\cite{Fu2022_JFM, grandmaison_brancherie_salsac_2021,
D2SM00179A,keller_skalak_1982, Finken08, Shaqfeh11}.

Molecular dynamics (MD) and Monte Carlo provide another avenue for
simulating the interaction between JPs, solvents, and
substrates~\cite{Brandner2019, Baniketal2021,
HongCacciutoLuijtenGranick2008, C9NR05885K}. These methods often use
pair-potentials to describe the interaction between JPs, for example by
prescribing angle dependent forces and torques that bring the
hydrophobic sides of two amphiphilic JPs into opposition. The HAP
formulation of the present work also has attractive, long-range forces
and torques, but they are instead derived from a boundary value problem
for the molecular structure of water~\cite{Ma77, GoHaKo94, ErLjCl89,
Lietal05, Israelachvili80}. Unlike in MD or Monte Carlo simulations, the
HAP interactions are nonadditive~\cite{Fu20}, so that the interactions
between a pair of JPs is affected by the presence of other particles. 

In this work, we take advantage of the flexibility of the HAP model to
examine the effects of varying the distribution of hydrophobicity on JP
surfaces. Such variation of the boundary condition on JP surfaces has
been realized experimentally by using chemicals to adjust the polarity
of the viscous solvent~\cite{Zarzaretal2015,
KirillovaMarschelkeSynytska2019, doi:10.1021/la503455h}. We show that a
simple tuning of the hydrophobic distribution leads to transitions from
unilamellar to multilamellar or striated superstructures of JPs.
Focusing on the fluid-structure interactions that correspond to such
transitions, we investigate the deformation of these novel structures in
background flows and map out their collective behavior away from
equilibrium. Looking forward, including other fields like electric
potential for JPs synthesized with charged
polymers~\cite{HongCacciutoLuijtenGranick2008, doi:10.1021/la503455h,
KangHonciuc2018}, is straightforward within the context of the boundary
integral representations~\cite{kohl-cor-che-vee22}, opening further
lines of investigation.

\section{Governing Equations: Hydrophobic Attraction Potential Mobility Problem} 
Following the formulation in \citet{Fu2022_JFM}, 
the governing equations are expressed as a system for the position and
orientation of a collection of rigid Janus particles. We first pose the
Stokes equations for the mobility problem giving the hydrodynamic
interactions for the particle suspension. The hydrophobic forces come
from solving a screened Laplace equation. Particle collisions are
avoided through a near-field, pair potential.

\subsection{Mobility problem}
The Janus particles are suspended in a viscous solvent. The particles
are disks of radius $c$, center $\aa_i$, and orientation $\theta_i$
relative to the horizontal axis, where $i = 1, \ldots, N_b$ and $N_b$ is
the number of particles. The domain $\Omega(t) \subset \mathbb{R}^2$ is
the solvent phase and $t$ is time. The boundary of $\Omega$ is
$\bd\Omega = \Gamma_1 \cup \cdots \cup \Gamma_{N_b}$, where $\Gamma_i$
is the boundary of Janus particle $i$. Assuming inertial terms are
negligible, we have 
\begin{alignat}{3}
\label{eq:stokes}
  -\mu \Delta \uu + \nabla p &= \mathbf{0}, && \xx \in \Omega, \\
\label{eq:stokesMass}
  \nabla\cdot \uu &= 0, \qquad && \xx \in \Omega, \\
\label{eq:stokesFarfield}
  \uu - \uu_\infty &\to \mathbf{0}, && |\xx| \to \infty,
\end{alignat}
where $\uu$ is the velocity of the solvent, $p$ is the pressure,
$\uu_\infty$ is the background flow velocity, and $\mu$ is the constant
viscosity. The solvent velocity satisfies the no-slip boundary condition
for a rigid body motion
\begin{align}
  \label{eq:rigid-vel}
  \uu(\xx) = \vv_i + \omega_i (\xx - \aa_i)^\perp, 
    \quad \xx \in \Gamma_i,
\end{align}
where $\vv_i$ is the translational velocity, $\omega_i$ is the angular
velocity, and $\langle x, y \rangle^{\perp} = \langle -y, x\rangle$. 

Without inertia, $(\uu, p)$ also satisfy the force-free and torque-free
conditions. This means that the force and torque from the hydrodynamic
stress on each particle balance the total force and torque coming from
the hydrophobic potential and repulsion. To compute this balance, we
define the free energy~\cite{Ma77, GoHaKo94, Fu20, Fu2022_JFM}
\begin{align}
\label{eq:free_energy}
F = \gamma
  \int_{\Omega} \left(\rho |\nabla u|^2 + \rho^{-1} u^2 \right)
\,d\xx
+ \frac{M}{2}
\sum_{j \neq i} 
P\left(\frac{|\aa_i - \aa_j|-2c}{\rho_0}\right),
\end{align}
where $u(\xx,t)$ is as an order parameter for the structure of water,
$\rho$ is a decay length, and $\gamma$ is an interfacial tension. 

The order parameter $u(\xx,t)$ is assumed to minimize the free energy
$F$. Its governing equations $u(\xx,t)$ are a screened Laplace equation
boundary value problem
\begin{gather}
  \label{eq:SL}
  -\rho^2 \Delta u + u = 0,\quad \xx \in \Omega, \\
  \label{eq:SLbc}
  u = g, \quad \xx \in \bd\Omega, \quad 
  u \rightarrow 0, \quad |\xx| \rightarrow \infty.
\end{gather}
The boundary condition $g$ encodes hydrophobic properties of the
particle-solvent interface.

The dimensionless repulsion profile $P$ takes the form $P(s) = 1 -
\sin(\pi s/2)$ for $0 \leq s < 1$ and $P(s) = 0$ for $1 \leq s <
\infty$. The parameter $\rho_0$ is an ``effective" repulsion distance between
particle surfaces, below which the steric repulsion between a pair of
JPs becomes important. The parameter $M$ is the repulsion modulus.

To calculate the force $\FF_i$ and torque $T_i$ on $\Gamma_i$ from the
free energy~\cite{Fu20}, we compute the variation of the domain of the
free energy~\eqref{eq:free_energy} subject
to~\eqref{eq:SL}--\eqref{eq:SLbc}, and obtain 
\begin{align}
  \label{eq:FandTdef}
  \FF_i = \int_{\Gamma_i} \mathbf{T}\nnu \, \dif s
  - \frac{M}{\rho_0}
  \sum_{j \neq i}
  \frac{\aa_i - \aa_j}{|\aa_i - \aa_j|}
P'\left(\frac{|\aa_i - \aa_j|-2c}{\rho_0}\right)
  ,\quad
T_i = \int_{\Gamma_i} (\xx - \aa_i)^{\perp}\cdot (\mathbf{T}\nnu) \, \dif s, 
\end{align}
for $i = 1,\ldots,N_b$ with particle outward normal $\nnu$.  The formulas
in \eqref{eq:FandTdef} involve the second-order hydrophobic stress tensor
\begin{align}
\mathbf{T} = \gamma \left[ \rho^{-1} u^2 \mathbf{I}
  + \rho \left(|\nabla u|^2 \mathbf{I} - 2\nabla u \nabla u^T\right)\right].
\end{align}
The repulsion is rotationally symmetric and
does not contribute to the torque $T_i$.

Since the force and torque from the hydrodynamic stress on each particle
balance the total force and torque coming from the hydrophobic potential
and repulsion, we have
\begin{equation}
  \label{eq:force}
 \int_{\Gamma_i} \ssigma \cdot \nnu \, \dif s = \FF_i, \quad
 \int_{\Gamma_i} (\xx - \aa_i)^\perp \cdot 
  (\ssigma \cdot \nnu) \, \dif s = T_i,
  \quad i=1,\ldots,N_b,
\end{equation}
where $\ssigma = -p \mathbf{I} + \mu \left(\nabla \uu + \nabla \uu^T
\right)$ is the hydrodynamic stress tensor.

To perform a single time step, we first solve for the HAP by
solving~\eqref{eq:SL}--\eqref{eq:SLbc}, and then compute the force and
torque~\eqref{eq:FandTdef}. Then the Stokes
equations~\eqref{eq:stokes}--\eqref{eq:stokesFarfield} are solved, and
the force-free and torque-free conditions are used to compute the rigid
body motion~\eqref{eq:rigid-vel}.
Once~\eqref{eq:stokes}--\eqref{eq:force} have been solved, the
velocities $(\vv_i, \omega_i)$ from~\eqref{eq:rigid-vel} are used to update the
particle positions and orientations.

\subsection{Boundary integral representations}
We recast~\eqref{eq:stokes}--\eqref{eq:rigid-vel}
and~\eqref{eq:SL}--\eqref{eq:SLbc} as boundary integral equations (BIEs)
and discretize each BIE at $N$ points on each of the $N_b$ particles
with a collocation method. To express the solution of
\eqref{eq:SL}--\eqref{eq:SLbc}, we adopt the double layer potential
\begin{align}
\label{eq:SL_BIE}
u({\xx}) = \DD[\sigma](\xx) = \int_\Gamma 
  \pderiv{G(\xx-\yy)}{\nnu_\yy}\sigma(\yy)\, \dif s_\yy, 
  \quad \xx \in \Omega,
\end{align}
where $G(\xx) = \frac{1}{2\pi}K_0(|\xx|/\rho)$ is the fundamental
solution to the screened Laplace equation~\eqref{eq:SL}, $K_0$ is the
zeroth-order modified Bessel function of the second kind, $\nnu_\yy$ is
the unit outward normal at $\yy$, and $\sigma$ is a scalar-valued
density function. The subscript in $\dif s_\yy$ denotes integration with
respect to $\yy \in \Gamma$. To satisfy the boundary
condition~\eqref{eq:SLbc}, the density function must satisfy\cite{Hsiao2008}
\begin{align}
  \label{eq:SL_BIE2}
  g(\xx) = \frac{1}{2} \sigma(\xx) + \DD[\sigma](\xx), \quad
    \xx \in \Gamma.
\end{align}

For the velocity, we use the representation
\begin{align}
  \label{eq:velocity}
  \uu(\xx) = \uu_\infty(\xx) + \DDD[\eeta](\xx) + 
    \sum_{i=1}^{N_b} \left(\SS(\xx,\aa_i) \cdot \FF_i + 
    \RR(\xx,\aa_i) T_i\right), \quad \xx \in \Omega,
\end{align}
where $\eeta$ is a vector-valued density function and
\begin{align}
  \DDD[\eeta](\xx) = \frac{1}{\pi} \int_{\Gamma} 
    \frac{(\xx - \yy) \cdot \nnu_\yy}{|\xx - \yy|^2}
    \frac{(\xx - \yy) \otimes (\xx - \yy)}{|\xx - \yy|^2}
    \cdot \eeta(\yy)\, \dif s_\yy.
\end{align}
The Stokeslets and Rotlets are
\begin{align}
  \SS(\xx,\aa_i) = \frac{1}{4\pi} \left(-\log |\rr|\mathbf{I} +
    \frac{\rr \otimes \rr}{|\rr|^2}\right), \quad 
  \RR(\xx,\aa_i) = \frac{1}{4\pi} \frac{\rr^\perp}{|\rr|^2}, 
\end{align}
respectively, where $\rr = \xx - \aa_i$~\cite{pow-mir1987}. Letting
$\xx$ approach $\Gamma_i$ in~\eqref{eq:velocity}, applying the jump
condition of the double layer potential~\cite{poz1992}, and imposing the
no-slip boundary condition~\eqref{eq:rigid-vel}, the density function
$\eeta$, translational velocity $\vv_i$, and rotational velocity
$\omega_i$ satisfy
\begin{alignat}{3}
  \label{eq:vel_BLM_rep}
  \vv_i + \omega_i (\xx - \aa_i)^\perp &= \uu_\infty(\xx)
    -\frac{1}{2} \eeta(\xx) + \DDD[\eeta](\xx) 
    + \sum_{j=1}^{N_b} 
    \left(\SS(\xx,\aa_j) \cdot \FF_j + \RR(\xx,\aa_j) T_j\right),\\
  \label{eq:vel_BLM_rep2}
  \int_{\Gamma_i} \eeta \, \dif s &= \mathbf{F}_i, \quad
  \int_{\Gamma_i} \eeta \cdot (\xx-\aa_i)^\perp \, \dif s = T_i,
\end{alignat}
for $\xx \in \Gamma_i$ and $i = 1,\ldots,N_b$.
Equation~\eqref{eq:vel_BLM_rep}--\eqref{eq:vel_BLM_rep2} does have a
unique solution, but we note that alternative full-rank layer potential
representations for rigid body motions are possible~\cite{rac-gre2016,
cor-gre-rac-vee2017}.

We discretize~\eqref{eq:SL_BIE2}
and~\eqref{eq:vel_BLM_rep}--\eqref{eq:vel_BLM_rep2} using high-order
interpolation-based quadrature rules. Integrals that are smooth are
computed with the spectrally-accurate trapezoid rule, and
nearly-singular integrals, caused by close contact between two
particles, are computed with a high-order interpolation-based quadrature
rule~\cite{qua-bir2014}.

After discretizing and applying quadrature, the result is an $NN_b
\times NN_b$ and $2NN_b \times 2NN_b$ linear system
for~\eqref{eq:SL_BIE2}
and~\eqref{eq:vel_BLM_rep}--\eqref{eq:vel_BLM_rep2}, respectively.
These are solved with matrix-free GMRES, and we guarantee that the
number of GMRES iterations is mesh-independent by using second-kind
BIEs. Once the translational and rotational velocity are computed, the
position and orientation of each JP is updated using the second-order
Adams-Bashforth scheme.


\section{Model parameters and boundary conditions} The interactions
generated by the HAP-mobility problem formulation
\eqref{eq:stokes}--\eqref{eq:force} lead to particle self-assembly for a
broad range of parameters and boundary conditions. Generally speaking,
the effective distance of the hydrophobic interaction is set by the
screening length $\rho$. Smaller values of the repulsion distance
$\rho_0$ decrease the distance where attraction and repulsion are in
balance and lead to more compact particle assemblies. The rate of
self-assembly is proportional to the interfacial tension $\gamma$, and
inversely proportional to solvent viscosity $\mu$; it is roughly
inversely proportional to screening length $\rho$ \cite{Fu20}.

To give our system physical units, we set the model parameters using
phospholipids as a characteristic amphiphile in water~\cite{Boal}. For
the solvent viscosity, we use $\mu = 1$~mPa~s for the viscosity of water
at room temperature. Pure lipid components give a range of interfacial
tensions $0.7$--$5.3$~pN~nm$^{-1}$~\cite{KUZMIN2005, Petelska2012,
Jackson2016, GarciaSaez}. As in our previous
works~\cite{Fu20,Fu2022_JFM}, we use $\gamma=4.1$~pN~nm$^{-1}$ since
this value gave good agreement with elastic moduli of lipid bilayers.
The particle radius $c = 1.25$~nm gives a diameter representative of
phospholipid length~\cite{Boal}, and the screening length $\rho = 5$~nm
derives from experimental force-distance measurements of hydrophobic
attraction~\cite{ErLjCl89,Lietal05,Israelachvili80,Jackson2016}. Based
on empirical studies, the repulsion modulus $M = 2$~pN and repulsion
distance $\rho_0 = 0.5$~nm give an interparticle distance around one
particle radius. This ensures the accuracy of the boundary integral
representations~\eqref{eq:SL_BIE} and~\eqref{eq:velocity} without having
to employ overly aggressive mesh refinement. Altogether, the above
parameter set gives a characteristic time $1$~ns and characteristic
length $1$~nm.

\subsection{Boundary conditions and equilibrium configurations}
The boundary condition $g(\xx)$ in
\eqref{eq:SLbc} defines the spatial distribution of
hydrophobicity and hydrophilicity.  On a hydrophobic region of the
particle surface, representing a hydrocarbon-water interface,
for example, $g(\xx)$ takes relatively large values.
On an apolar, hydrophilic region, $g(\xx)$ is close to zero.

In the present work, we consider boundary conditions of the form
\begin{align}
  \label{eq:bc-type}
g(\xx) = a(b + \cos \alpha),\quad a = (\pi c(2b^2 + 1))^{-1/2},\quad \xx \in \Gamma_i,
\end{align}
where $\alpha$ is the angle between the vector $\xx - \aa_i$ and the
particle director $\dd_i = (\cos \theta_i, \sin \theta_i)$.  The
parameter $a$ normalizes the boundary conditions so that
$\int_{\Gamma_i} g^2(\xx) \, \dif s = 1$.  The motivation for this
normalization is that the hydrophobic attraction part of the free energy
converges to the integral of the square of the boundary data in the
zero-screening length limit. We refer to the side of the particle where
$\alpha = 0$ as the tail and the side where $\alpha = \pi$ as the head.

\begin{figure}[h!]
\begin{center}
  \includegraphics[width=\textwidth]{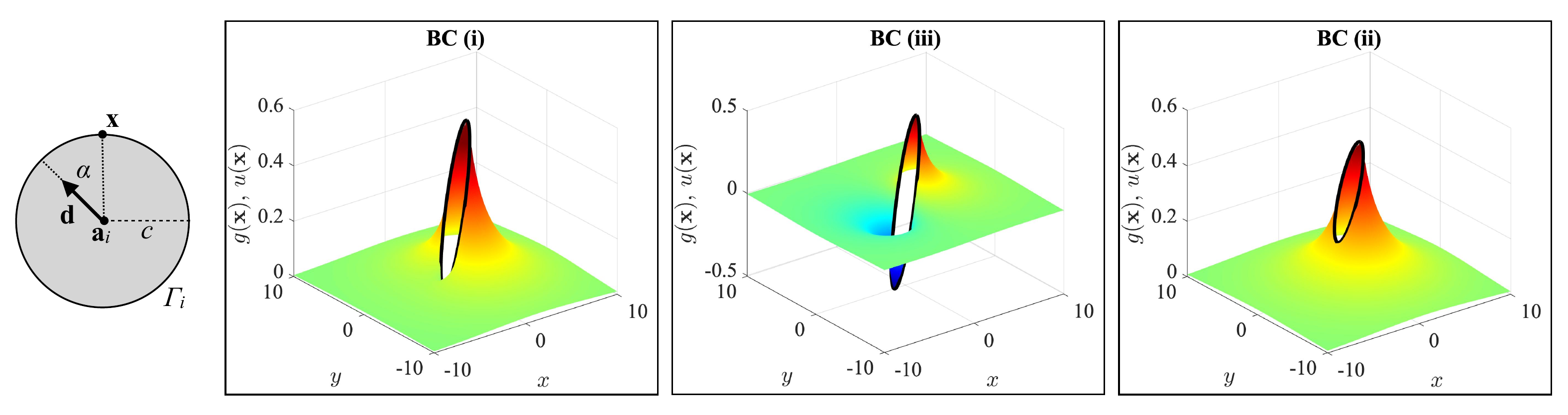}
\end{center}
\begin{caption}{\label{fig:boundary_conditions} The leftmost diagram
  illustrates the particle $\Gamma_i$ with center $\aa_i$, radius $c$,
  and director $\dd$, along with the angle $\alpha$ used to define
  $g(\xx)$ in~\eqref{eq:bc-type}. The three right plots show $g(\xx)$
  (black curve) for BC (i), (ii), and (iii) when $\dd = (1,0)$. The
  surfaces are the respective solutions $u(\xx)$ of~\eqref{eq:SL} for a
  single, isolated particle.}
\end{caption}
\end{figure}

\begin{figure}[h!]
\begin{center}
  \includegraphics[height=0.3\textheight]{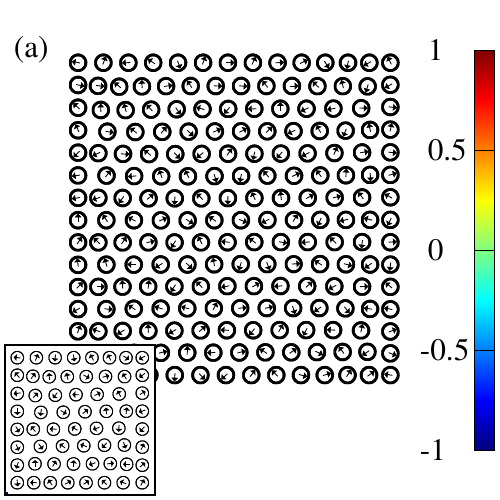}
  \includegraphics[height=0.3\textheight]{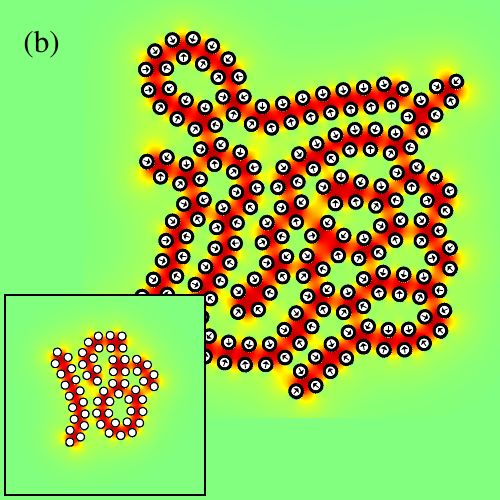}\\
  \includegraphics[height=0.3\textheight]{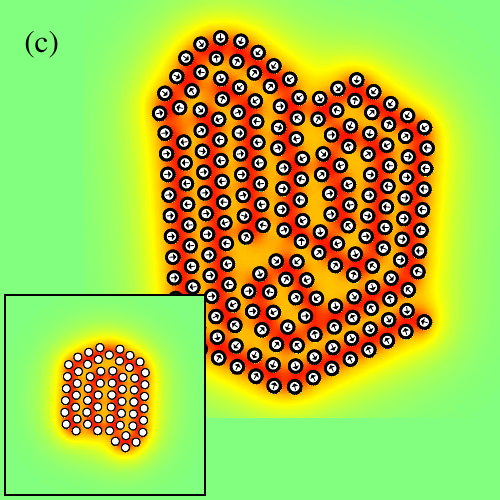}
  \includegraphics[height=0.3\textheight]{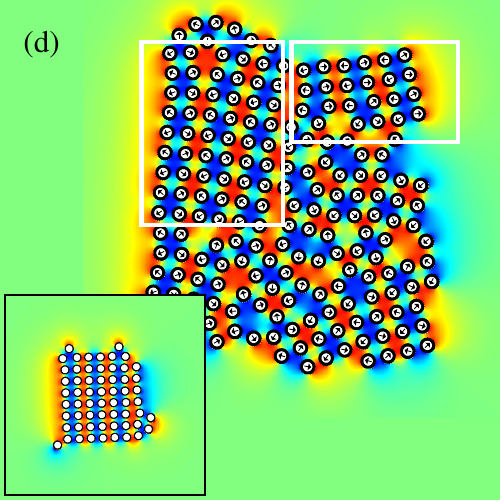}
\end{center}
\begin{caption}{\label{fig:relax}
  Panel (a) shows $N_b = 198$ particles confined to a square with random
  positions and orientations and the inset shows a $N_b=60$ particles
  configuration. The particles self-assemble into distinct
  configurations, depending on the boundary
  condition~\eqref{eq:bc-type}. Panel (b) is for $b=1$ and the boundary
  condition mimics that of an amphiphilic particle giving bilayer
  structures. Panel (c) is for $b=2$ which gives rise to a
  multilamellar configuration. In panel (d), the case $b = 0$ is for a
  water structure with positive/negative charge. The false color maps
  use blue for $u < 0$, green for $u = 0$, and yellow then red for $u >
  0$. The inset in panels (b)--(d) are configurations using $N_b=60$ and
  share the same color map as the main panels.}
\end{caption}
\end{figure}

We set $\max g>0$ to guarantee at least partial hydrophobicity of the JP
surface. Then, the boundary condition (BC) can be classified into three
categories: $\min g(\xx) = 0$, $\min g(\xx) >0$, and $\min g(\xx) <0$
where the extrema are taken for $\xx \in \Gamma_i$. These three
categories correspond to three characteristic values of $b$ in
\eqref{eq:bc-type}: $b=1$, $b=2$ and $b=0$, respectively. 
To see the effect of the shift parameter $b$, we simulate the dynamics
of 198 particles and 60 particles in a quiescent flow as shown in
Figure~\ref{fig:relax}. Three distinct configurations emerge: (i)
bilayer (amphiphilic) for $b = 1$, (ii) multilamellar (biased hydrophobic) for $b = 2$, and (iii) striated (bipolar)
for $b = 0$. Supplementary Movie S1 shows the self-assembly process for
the three cases. In the following,
the terms ``multilamellar" and ``biased hydrophobic"
are used  interchangeably to refer to BC (ii), while
``striated" and ``bipolar" are used interchangeably to refer to BC (iii).

Figure~\ref{fig:relax}(a) shows the initial configurations with
orientations of the 198-particle and 60-particle cases, respectively.
BC (i) with $b = 1$ mimics an amphiphilic particle. The boundary data
$g$ is everywhere nonnegative, and takes the maximum value $a$ on the
hydrophobic tail (red) the value $0$ on the hydrophilic head (green).
The interaction is attractive, and particles collectively orient their
hydrophobic sides to form bilayers (Figure~\ref{fig:relax}(b)). The
inset shows the equilibrium state for the 60-particle run. Both
$N_b = 60$ and $N_b = 198$ cases
form bilayers with multiple domains.

In BC (ii) with $b = 2$, the boundary data is everywhere positive. It
takes the maximum value $3a$ on the tail (red) and the minimum value $a$
on the head side (yellow). Both sides of the Janus particle are
hydrophobic but more so on the $\alpha = 0$ side. Over a short time
scale, these particles also self-assemble into bilayers. But unlike for
BC (i), the heads are also hydrophobic and so in the long-time dynamics
these bilayers form multilamella structures~\cite{C9NR05885K}. The
number of layers depends primarily on the number of particles.
Figure~\ref{fig:relax}(c), for example, shows a multilamellar structure
with four layers and one with two layers in the inset.

Finally, BC (iii) with $b=0$ corresponds to a water structure with
positive/negative charge~\cite{MaRa76, Ma77}. Here, the head repels the
tail of other particles and the particles initially form chains with
their directors perpendicular to the length of the chain. The chains
form stria where the particles lie on a square grid and the orientations
alternate between layers (Figure~\ref{fig:relax}(d)).

\begin{figure}
  \begin{center}
\includegraphics[width=1.0\textwidth]{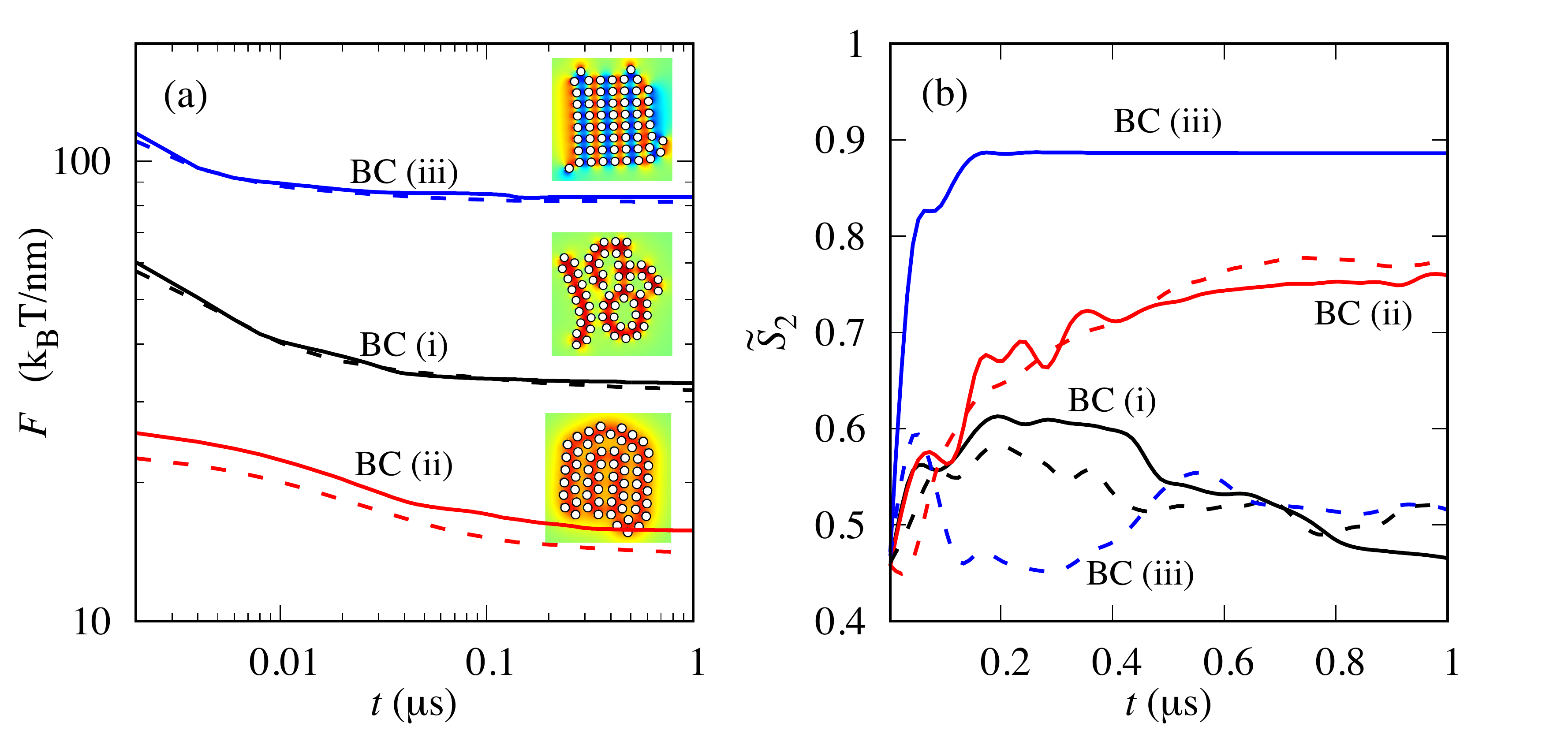}
  \end{center}
  \vspace{-20pt}
  \caption{\label{fig:relax_energy}
  Panel (a) shows that the free energy~\eqref{eq:free_energy} decreases
  in the mobility problem formulation. The initial energies are for the
  configuration in Figure~\ref{fig:relax}(a) and the final energies are
  for the near-equilibrium configurations shown in
  Figures~\ref{fig:relax}(b)--(d). The solid curves are the energies for
  the $N_b=60$ cases and the dashed curves are the energies for the $N_b
  = 198$ cases normalized by $60/198$, showing that the energy
  approximately scales with the number of particles.  Panel (b) shows
  the orientational parameters $\tilde{S}_2$ for all 6 relaxation runs.
  The curve styles are identical those in panel (a). The
  near-equilibrium configurations depend on the initial configurations
  and the results show that for both $N_b=60$ and $N_b=198$ cases,
  bilayer and multi-lamellar configurations have similar trends in
  orientational parameters. The results of the striated configuration,
  however, depend on the initial geometrical setup.}
\end{figure}

\section{Measuring deformation}
To quantify the collective hydrodynamics of JPs,
we use the free energy $F$, a strain parameter $E$, and two scalar order
parameters $S_{2}$ and $\tilde S_2$ to measure the deformation of
particle configurations under background flows. 

First we simplify the form of the free energy~\eqref{eq:free_energy}.
Using integration by parts and~\eqref{eq:SL}, we obtain
\begin{equation}
\label{eq:free_energy2}
F = -\gamma
\int_{\Gamma} \rho g \nabla u \cdot \nnu \,\dif s
+ \frac{M}{2}
\sum_{j \neq i} 
P\left(\frac{|\aa_i - \aa_j|-2c}{\rho_0}\right).
\end{equation}
Here, we have substituted $g$ for $u$ since the boundary values are
given. However, evaluating $\nabla u \cdot \nnu$ on $\Gamma$ based
on~\eqref{eq:SL_BIE} involves calculating a quadruple layer potential
which has a well-known obstacle in numerical implementation.  To
overcome this obstacle, we use
\begin{equation}
\label{eq:normal_deriv}
\nabla u(\xx) \cdot \nnu(\xx)=
-\frac{1}{\rho^2} {\bf t}_\xx\cdot \SSS[\sigma{\bf t}](\xx)
+ \frac{\dif}{\dif s}\SSS\left[\frac{\dif \sigma}{\dif s}\right](\xx), \quad \xx \in \Gamma.
\end{equation}
Here, $\tt(\xx)$ is the tangent vector and $\dif/\dif s$ is the
arclength derivative. See the Appendix in \S\ref{sec:appendix} for the
proof and details. Substituting~\eqref{eq:normal_deriv}
into~\eqref{eq:free_energy2} for the normal derivative leads to the
single layer, $\SS$, which is more straightforward to evaluate and we
compute the arclength derivative using spectrally accurate Fourier
transform. We have validated the form of~\eqref{eq:normal_deriv} by
extrapolating the primal energy~\eqref{eq:free_energy} along test curves
placed slightly outside the particles.

Figure~\ref{fig:relax_energy}(a) tracks the free energy profiles for all
relaxation runs.  For the 198-particle cases, we normalized the energies
by 60/198 and the results show that all energies decrease with good
agreements in all boundary conditions (dashed curves).  This can be
considered as evidence that the free energy per particle with specified
boundary conditions is independent of the total particle number $(N_b)$.

To measure positional order, we use a number, which we call the strain
parameter, 
\begin{equation}
\label{eq:SP}
E = \frac{1}{N_b} \sum_{i=1}^{N_b}
\left\|\frac{1}{2}(\mathsf{F}_i^T \mathsf{F}_i - I)\right\|,
\end{equation}
where $\mathsf{F}_i$ is an approximate deformation gradient at particle
$i$. The argument inside the Frobenius norm $\| \cdot \|$ is the
Green-Lagrange strain tensor. The motivation for~\eqref{eq:SP} is as
follows. For relatively weak background flow strengths, the
near-equilibrium particle configurations behave as an elastic solid.
There is a map $\ff(\xx,t)$ from the reference (equilibrium)
configuration with $\ff(\aa_i(0),t) = \aa_i(t)$ for $i = 1,\ldots,N_b$.
Using linear approximation,
\begin{align}
\aa_j(t) - \aa_i(t) = \ff(\aa_j(0),t) - \ff(\aa_i(0),t)
\approx \nabla \ff(\aa_i(0),t)(\aa_j(0) - \aa_i(0)).
\end{align}
To approximate the deformation gradient $\nabla \ff(\aa_i(0),t)$ we
solve the overdetermined system 
\begin{align}
\aa_j(t) - \aa_i(t) = \mathsf{F}_i(\aa_j(0) - \aa_i(0)),\quad j =
  1,\ldots, N_b,
\end{align}
for $\mathsf{F}_i$ by weighted least squares. The weights $w_i =
\exp(-\|\aa_j(0) - \aa_i(0)\|/4c)$ with particle radius $c$ ensure that
the linear approximation is valid for particles near $\aa_i$.

Finally, we use the scalar order parameter $S_2$ to quantify the orientational order\cite{Selinger2016}:
\begin{equation}
  \label{eq:S2}
S_2 = \frac{1}{N_b} \sum_{i=1}^{N_b} \frac{1}{2}(3\cos^2(\theta_i - \bar \theta) - 1).
\end{equation}
Here $\bar \theta$ is a circular mean defined as the orientation of the
principal eigenvector of the matrix $\dd_1\dd_1^\top + \cdots +
\dd_{N_b}\dd_{N_b}^\top$. Defined as such, $S_2$ is always in the range $0 \le S_2 \le 1$.
A value $S_2 = 1$ indicates that all
particle directors lie on a common axis e.g., two parallel vectors with
possibly opposite direction are ordered. A value below 1 indicates imperfect order, and $S_2=0$ indicates
complete disorder.

We modify $S_2$ to account for the bilayer and multilamellar structures.
In these cases, the directors are more or less uniformly distributed
along a circle, even though there is orientational order between
neighboring particles. To account for local order, we instead use
\begin{equation}
  \label{eq:localS2}
\tilde{S}_2(t) = \frac{1}{N_b} \sum_{i=1}^{N_b}
\frac{1}{2}(3\cos^2(\theta_i - \bar \theta_i) - 1),
\end{equation}
where $\bar \theta_i$ is the circular mean restricted to particles
indexed by $j$ with $\|\aa_i - \aa_j\| < 4c$.
This cutoff distance was chosen empirically
so that the average includes nearest neighbors
but excludes next nearest neighbors.
In practice, if a particle
is isolated and has no neighbors within a distance $4c$, then we exclude
it from the sum~\eqref{eq:localS2}. 

Figure~\ref{fig:relax_energy}(b) tracks the orientational order
parameter $\tilde{S}_2$. Both bilayer and multilamellar structures show
similar trends in the small and large particle number systems
(Figure~\ref{fig:relax_energy}(b) black curves and red curves,
respectively). The multilamellar structures are highly ordered in all
cases (Figure~\ref{fig:relax}(c)) whereas the bilayer structures are
disordered because they consist of several components forming isolated
bilayers, micelles, and vesicles (Figure~\ref{fig:relax}(b)).

The aggregation for striated configurations forms more than one pattern
when the number of particles is varied (Figure~\ref{fig:relax}(d)). We
observe that when the number of particles is fewer, there is only a
single pattern in particle orientations where the directors are more or
less parallel and alternate directions (Figure~\ref{fig:relax}(d),
inset). Larger number of particles results in two different patterns:
the alternating sign pattern as in the small particle number case
(Figure~\ref{fig:relax}(d), top right rectangle) and one where the
directors reflect across the lines parallel to the stria and across the
lines perpendicular to the stria (Figure~\ref{fig:relax}(d), top middle
rectangle). As a result, there are different trends in orientational
order (Figure~\ref{fig:relax_energy}(b), solid and dashed blue curves,
respectively).

In the Results section, the equilibrium configurations for BCs (i),
(ii), and (iii) (Figure~\ref{fig:relax}, insets) are used as initial
data in the background flow simulations. For the bilayer case BC (i),
we include an alternate initial condition consisting of a single,
circular vesicle (c.f., Figure~\ref{fig:Ves_shear}).


\begin{figure}
  \begin{center}
   \includegraphics[width=1.0\textwidth]{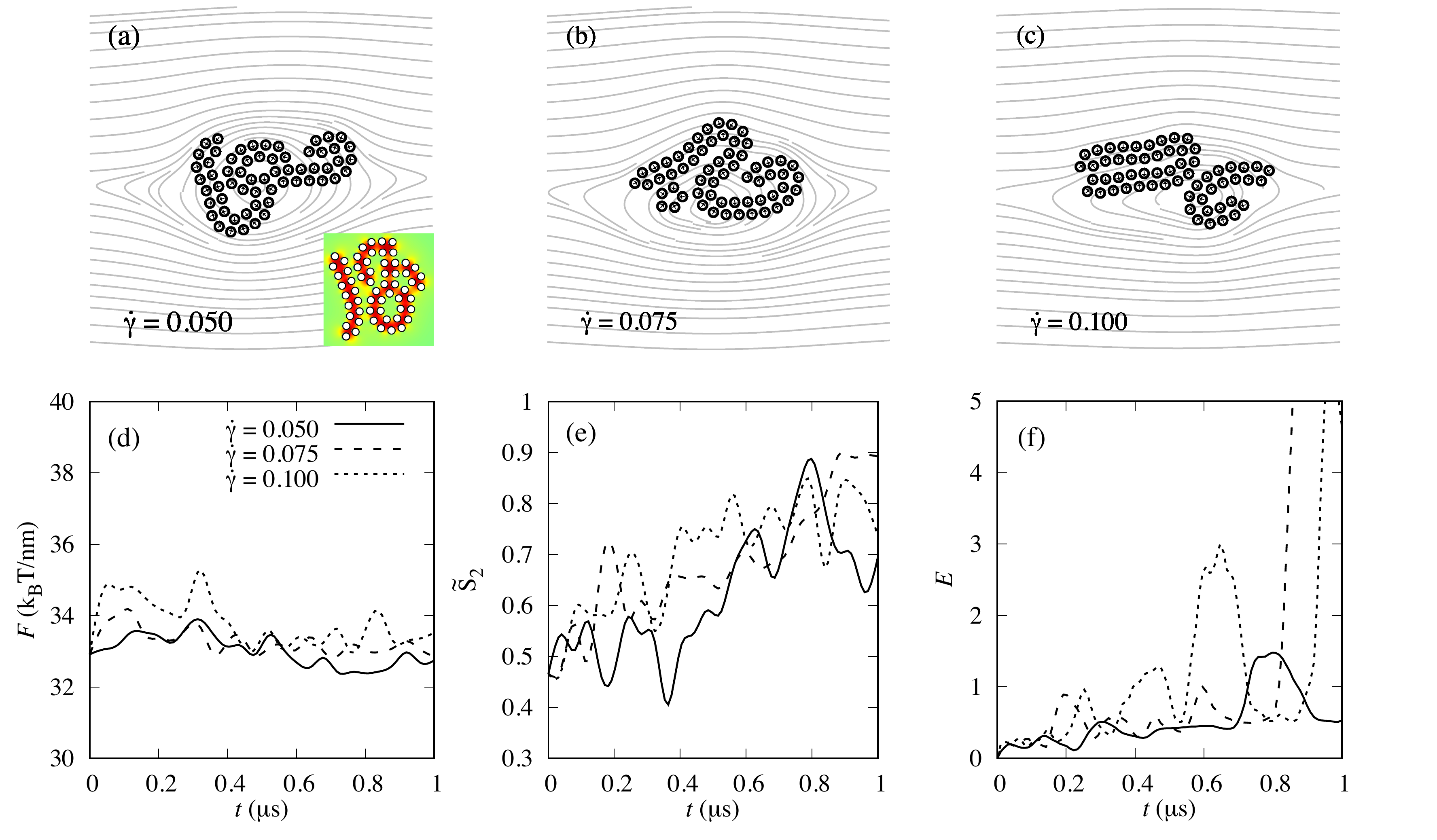}
  \end{center}
  \caption{
    \label{fig:BC1_shear}
    Panels (a)--(c) are a snapshot at $t=0.5\ \upmu$s of the
    multiple-component bilayer with shear rate $\dot \gamma = \{0.05,
    0.075, 0.1\}$. The initial configuration is shown in inset of panel
    (a). The streamlines are plotted in the background. Panel (d) shows
    the free energies, panel (e) shows the orientational parameter
    $\tilde{S}_2$, and panel (f) shows positional parameter E.}
\end{figure}

\section{Results}
\label{sec:results}
In this section, we subject the JP structures to various
background flows. The purpose is to understand how the self-assemblies
respond to external forcing. We then compare effective
properties of materials comprised of JPs as a function of
their amphiphilic properties.

\subsection{Bilayer configurations (BC (i)) in a linear shear flow}
To begin, we place the bilayer structure (with BC (i)) from the inset of
Figure~\ref{fig:relax}(b) in a linear shear flow
\begin{equation}
\uu_\infty(\xx) = \dot\gamma \text{ ns }^{-1} ({\bf e}_y \cdot \mathbf{x}) {\bf e}_x,
\end{equation}
where $\dot\gamma$ is the dimensionless shear rate, and ${\ee}_x$ and
${\ee}_y$ are horizontal and vertical unit vectors, respectively. The
initial bilayer structure consists of a small self-enclosing bilayer
(vesicle) and several pieces of bilayers.
Figure~\ref{fig:BC1_shear}(a)--(c) shows a snapshot at $t = 0.5$
$\upmu$s of the bilayer for different shear rates $\dot\gamma=
0.05,0.075,0.01$. The dynamics of the bilayer can be found in
Supplementary Movie S2, which shows
that the initial pieces of bilayers
merge and form a worm-like shape under a linear shear flow. At a low
shear rate, the bilayer constantly goes through rotation and extensional
deformation with the whole bilayer remaining intact
(Figure~\ref{fig:BC1_shear}(a)). As the shear rate increases to a
moderate value (Figure~\ref{fig:BC1_shear}(b)), the bilayer is observed
to shed a small piece of bilayer, similar to the asymmetric breakup of a
viscous drop in confinement \cite{DuFuZhuMaLi2016_AICHEJ}.
In Figure~\ref{fig:BC1_shear}(c) the high shear rate gives rise to two
separate, nearly equal bilayers, similar to the symmetric drop breakup
under a linear shear flow \cite{Stone1994_ARFM}. For the dynamic evolution, see
Supplementary Movie S4.


We observe that the order parameter $\tilde S_2$ in
Figure~\ref{fig:BC1_shear}(e) is on average increasing. In the initial
configuration, the structure has three or four pieces of bilayer.
Under the shear flow, the shedding and merging of bilayers results in an
overall decrease in distinct bilayer structures. Similarly, the free
energies are slightly decreasing (Figure~\ref{fig:BC1_shear}(d)), with
the overall changes occurring in the range of a few $\mathrm{k_BT}$/nm.
This suggests that the shear flow changes the energy landscape to move
the local equilibrium obtained from a random initial condition to an
alternate local equilibrium with less free energy and greater
orientational order  (Supplementary Movie S2).  Note that the free energy $F$ in
Figure~\ref{fig:BC1_shear}(d) has the unit of force ($\mathrm{k_BT}$/nm $\approx$ 4.114 pN)
because our system
is two dimensional.
%
%
Finally, the strain parameter diverges in the large shear rate cases
(Figure~\ref{fig:BC1_shear}(f))
because the various pieces of bilayer eventually completely separate and move away from one another
under the shear flow.

\begin{figure}
  \begin{center}
    \includegraphics[width=1.0\textwidth]{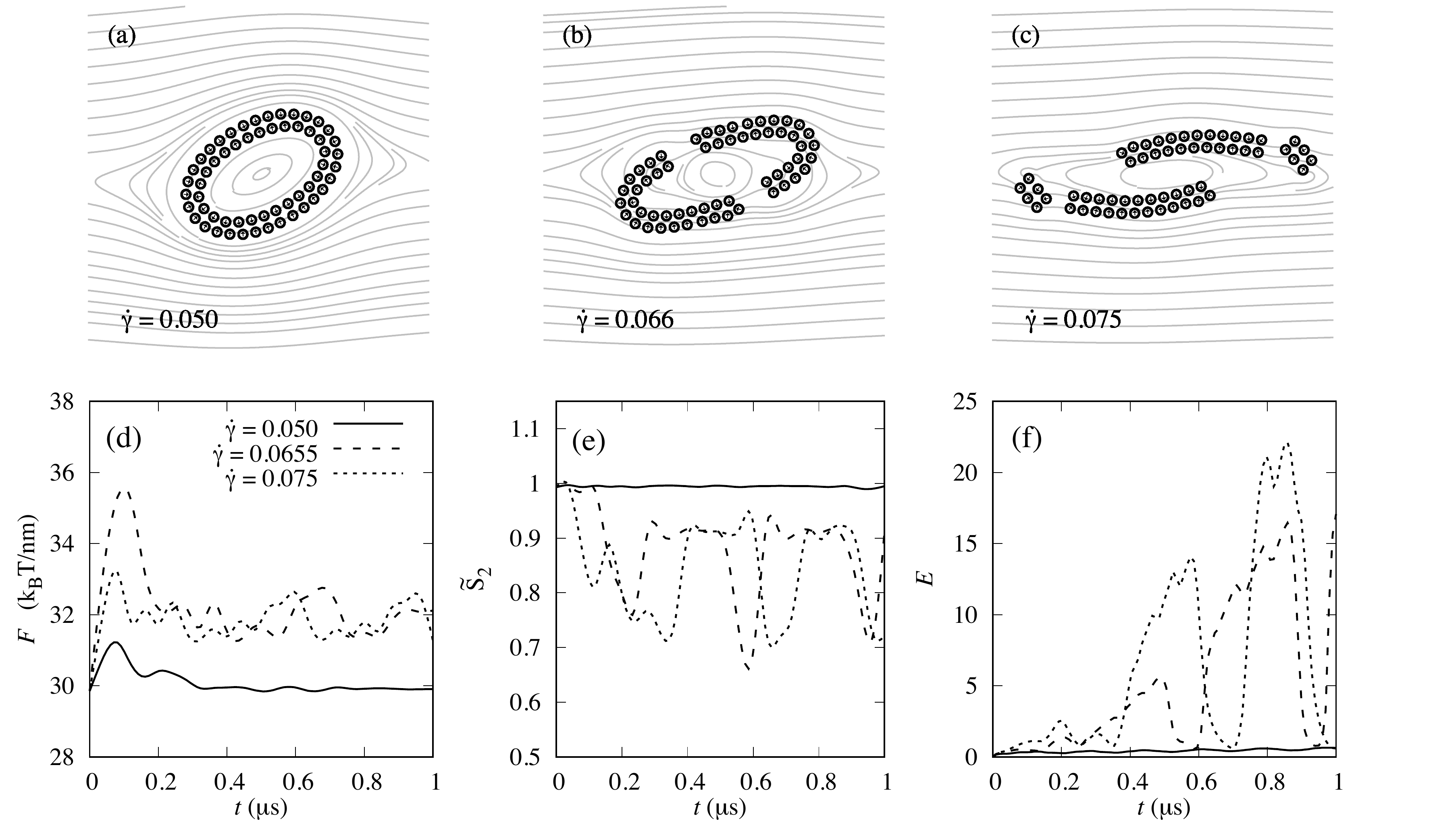}
  \end{center}
  \caption{
    \label{fig:Ves_shear}
A single vesicle in a shear flow. Panels (a)--(c) are snapshots for
  $\dot \gamma = \{0.05, 0.0655, 0.075\}$ at $t=0.16\ \upmu$s where the
  pre-relaxed initial configuration is shown in inset of panel (a). The
  streamlines are plotted in the background. Panel (d) shows the free
  energies; panel (e) shows orientational parameter $\tilde{S}_2$; panel
  (f) shows positional parameter E.}
\end{figure}
\citet{Fu2022_JFM} studied the hydrodynamics of a JP vesicle under a
linear shear flow. Focusing on drawing analogies to hydrodynamics of a
vesicle in continuum modeling (such as
tank-treading~\cite{keller_skalak_1982,Finken08,Shaqfeh11}, bilayer
slip~\cite{sch-vla-mik2010,denOtter2007,Zgorski2019}, and
permeability~\cite{chabanon2017, qua-gan-you2021}), our previous work
demonstrated the existence of a critical shear rate above which the JP vesicle
ruptures~\cite{grandmaison_brancherie_salsac_2021,D2SM00179A}. Based on
the these early results, we reinitialized the BC (i) configuration in
the form of a single, circular vesicle (rather than several bilayer
components) and illustrate how the three measures of structural
deformation correlate to the dynamics of a JP vesicle.

At shear rates $\dot\gamma=0.05, 0.066, 0.075$, the vesicle undergoes
familiar hydrodynamics, such as elongation along the extensional axis
and tank-treading motion, as shown in
Figure~\ref{fig:Ves_shear}(a)--(c)
and Supplementary Movie S2.
The circular shape is also a local equilibrium, with somewhat less
energy than the disordered state c.f., Figure~\ref{fig:BC1_shear}(d) and
Figure~\ref{fig:Ves_shear}(d). But unlike the dynamics of bilayers in
Figure~\ref{fig:BC1_shear}, the energies of the JP vesicle in
Figure~\ref{fig:Ves_shear}(d) jump from the baseline value 30
$\mathrm{k_BT}$/nm to about 32 $\mathrm{k_BT}$/nm, signaling the rupture
of the vesicle into disconnected bilayers (see
Figure~\ref{fig:Ves_shear}(d), long and short dashed curves).  After the
rupture, each piece circles around the vorticity at the center without
reconnection while the energies fluctuate slightly around a constant.

For the lowest shear rate case, the vesicle has nearly constant
orientational order $\tilde S_2 = 1$ (see Figure~\ref{fig:Ves_shear}(e),
solid curve). After the JP vesicle ruptures with $\dot\gamma= 0.066$ and
$\dot \gamma= 0.075$, we observe a greater variation in the order
parameter $\tilde{S}_2$ with a baseline value around $0.7$ (see
Figure~\ref{fig:Ves_shear}(e), dashed curves). The oscillations occur
due to the orbit of the two bilayer components, and $\tilde{S}_2$
increases to about $0.9$ when the components slide past one another and
are nearly parallel.
As in the previous case, peaks in the strain parameter $E$ correlate
well with changes in structural topology in the bilayer dynamics (dashed
curves in Figure~\ref{fig:Ves_shear}(f)).

%
\begin{figure}
  \begin{center}
    \includegraphics[width=1.0\textwidth]{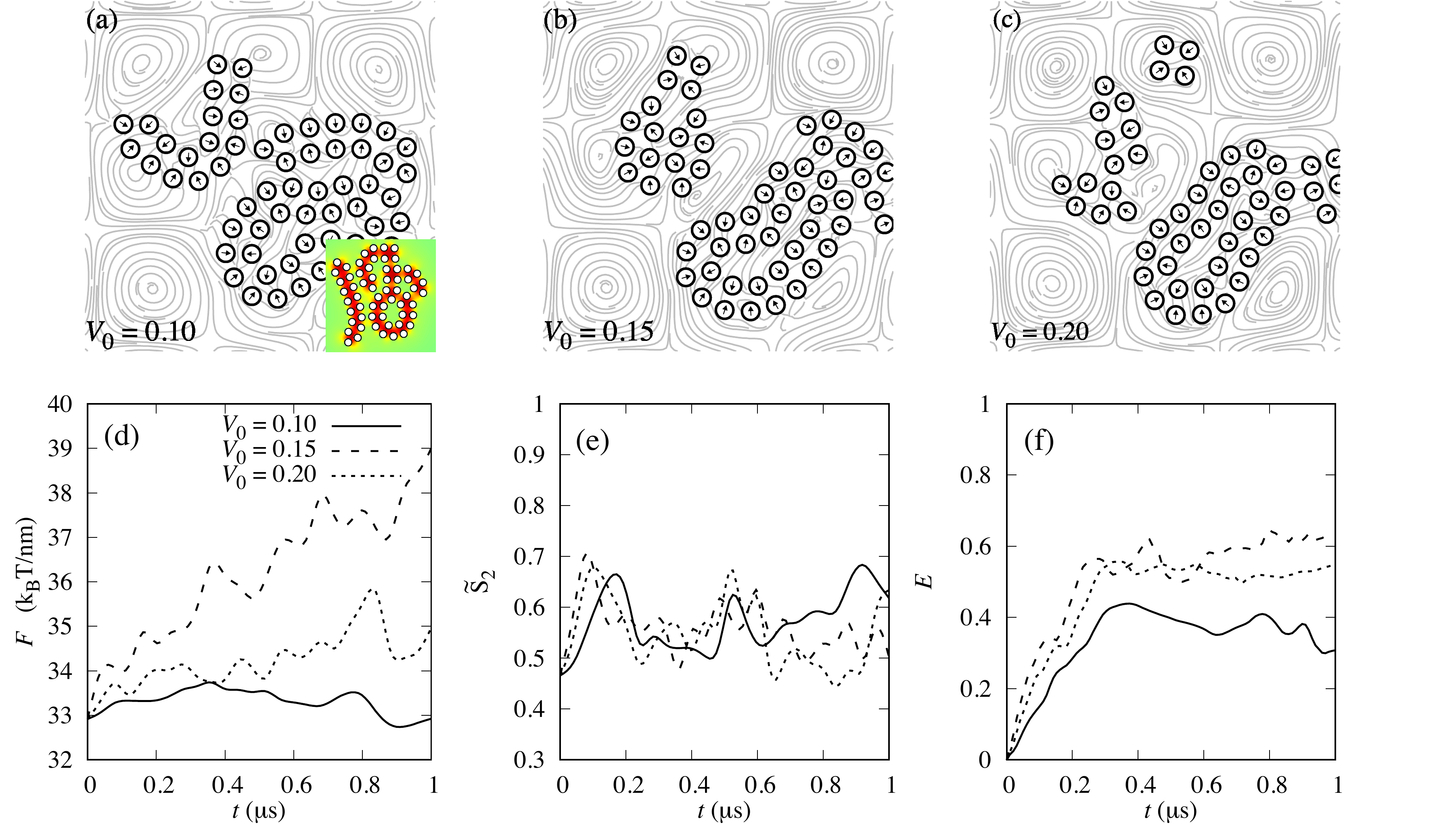}    
  \end{center}
  \vspace{-20pt}  
  \caption{\label{fig:BC1_TG} Bilayers with multiple domains in a
  Taylor-Green when $V_0 = \{0.1, 0.15, 0.2\}$ at $t=0.2\ \upmu$s. The
  pre-relaxed initial configuration is shown in the inset of panel (a).
  The streamlines are plotted in the background. Panel (d) shows the
  free energies; panel (e) shows the orientational parameter
  $\tilde{S}_2$; panel (f) shows the positional parameter $E$.}
\end{figure}
\subsection{Bilayer and vesicle configurations in a Taylor-Green flow}
Next, we subject the JP structures to a steady Taylor-Green (TG) background
flow
\begin{equation}
\label{eq:tg_flow}
\uu_\infty(\xx) = V_0\; \frac{\text{nm}}{\text{ns}}\;
\left(
-\cos\left(x/\lambda\right)
 \sin\left(y/\lambda\right)
         {\bf e}_x
         +\sin\left(x/\lambda\right)
         \cos\left(y/\lambda\right)
             {\bf e}_y\right),
\end{equation}
where $x = {\bf e}_x \cdot \xx$ and $y = {\bf e}_y \cdot \xx$
are the horizontal and vertical coordinates, respectively. 
The TG flow is a confining flow consisting of a checkerboard pattern
of cells with alternating circulation. We control the flow by the
dimensionless flow strength $V_0$ and the dimensionless cell size $l$
defining $\lambda = l$ nm. For most of the results in this subsection,
we use the parameter $l=2$.

\begin{figure}
  \begin{center}
    \includegraphics[width=1.0\textwidth]{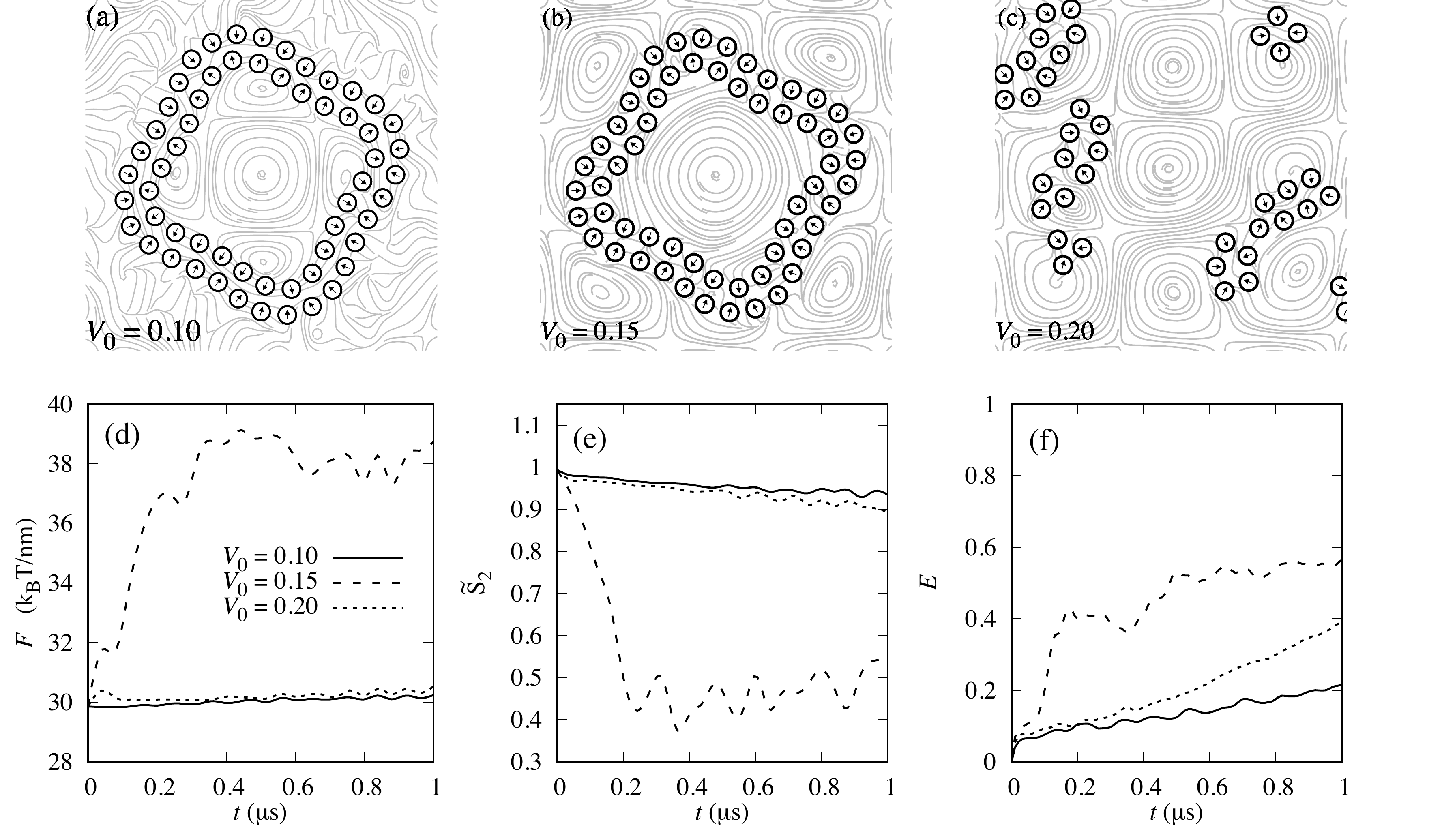}        
  \end{center}
\caption{\label{fig:ves_TG} A single vesicle in a TG flow. Panels
  (a)-(c) are snapshots for $V_0=\{0.1, 0.15, 0.2\}$ at $t=0.6\ \upmu$s
  where the pre-relaxed initial configuration is shown in inset of panel
  (a). The streamlines are plotted in the background. Panel (d) shows
  the free energies; panel (e) shows orientational parameter
  $\tilde{S}_2$; panel (f) shows positional parameter E.}
\end{figure}

There is no appreciable pattern in the deformation for the
case of bilayers with multiple pieces of bilayer in TG flow
(Supplementary Movie S3, right panel).
Figure~\ref{fig:BC1_TG}(a)--(c)
are the snapshots when $V_0=0.1,0.15,0.2$ at $t = 0.2$ $\upmu$s. At all
flow strengths, the bilayer pieces are separated and mixed by the
flow in neighboring cells. In the $V_0 = 0.2$ case, the flow breaks off
bilayers consisting of only a few particles, leading an increase in
energy (Figure~\ref{fig:BC1_TG}(d), long dashed curve). There is no
apparent structure to the change in the orientational order. There is an initial
increase in the strain parameter due to merging and rupture of bilayers, but the
strains remains bounded since the particles stay confined to the
neighbors of the central cell.

Patterns in the deformation, however, do emerge in the case of a vesicle
in TG flow (Supplementary Movie S3, left panel). For a vesicle in a Taylor-Green flow,
Figure~\ref{fig:ves_TG}(a)--(c) are configurations at $t=0.2\ \upmu$s
where $V_0=0.1,0.15,0.2$. Here, the circular vesicle takes on a steady
square shape for moderate flow rates $V_0 = 0.1$ and $V_0 = 0.15$. In
these cases, there is little change in the vesicle free energy and order
(Figure~\ref{fig:ves_TG}(d)--(f)). The vesicle disintegrates for $V_0 =
0.2$ (Supplementary Movie S5, left panel), leading to a jump in energy, drop in order, and increase in
strain, suggesting that the vesicle has a critical TG flow strength in
the interval $0.15 < V_0 <  0.2$ separating the intact and ruptured
end-states.

\begin{figure}
  \begin{center}
\includegraphics[width=1.0\textwidth]{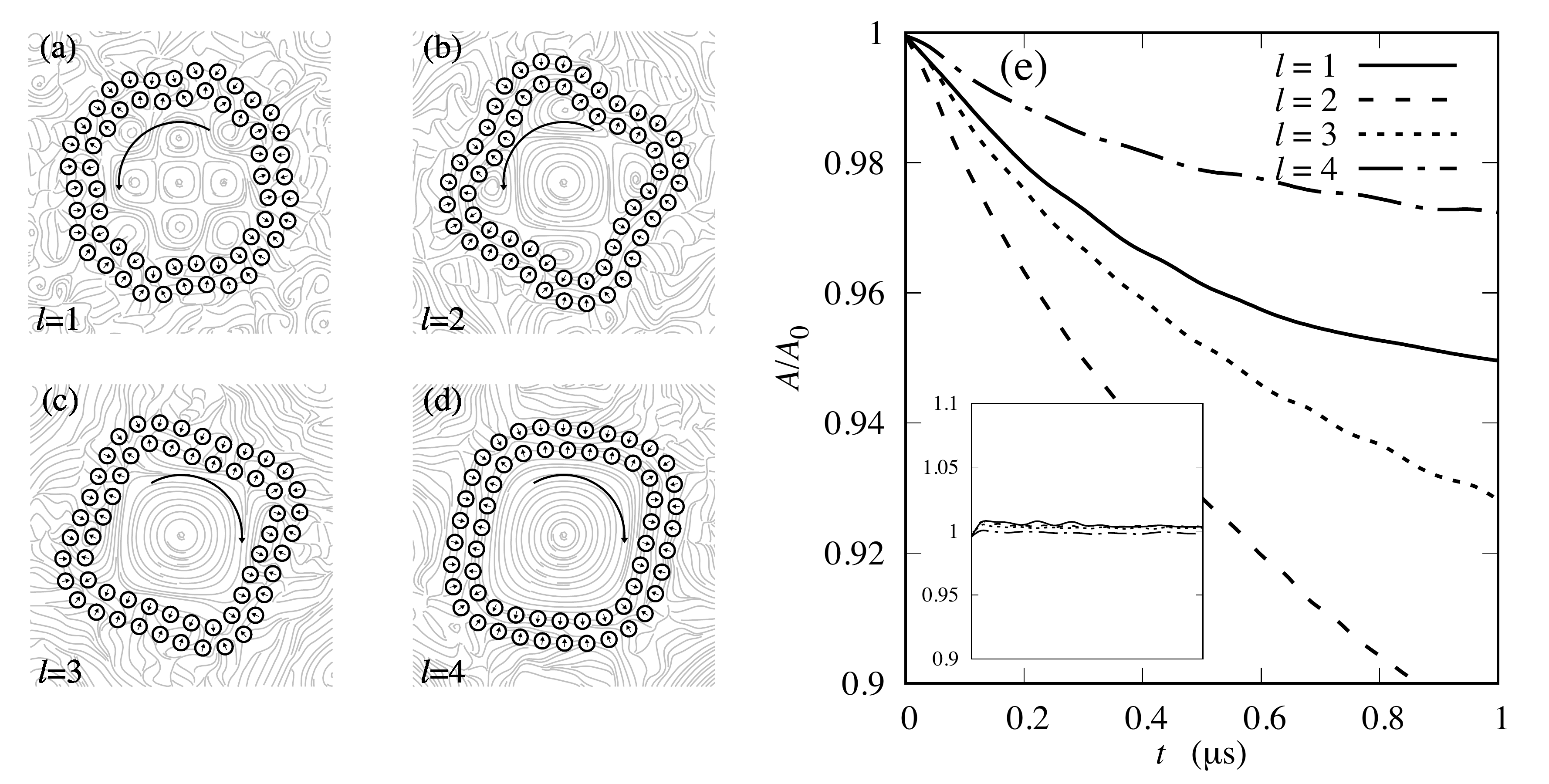}
  \end{center}
  \vspace{-20pt}  
  \caption{\label{fig:BTG_Scale} With $V_0=0.1$, panels (a)--(d) show
  the configurations with streamlines of a vesicle is in a TG flow at
  $t=1$ $\upmu$s with $l= 1,2,3,4$, respectively. In panel (e), the
  relative enclosed area $A/A_0$ calculated from the bilayer midplane
  decreases over time. The inset shows the relative length $l/l_0$ which
  remains nearly constant with small magnitude of oscillations. The
  inset and main panel have the same horizontal axis.}
\end{figure}

The vesicle shape is also related to the cell size $l$.
Figure~\ref{fig:BTG_Scale}(a)--(d) show the deformation of a vesicle in
a TG flow for $V_0=0.1$ when $t = 1$ $\upmu$s. For the smallest cell
size $l = 1$ tested, the vesicle is somewhat octagonal and passes
through multiple cells (Figure~\ref{fig:BTG_Scale}(a)). For the largest
cell size $l = 4$, the vesicle is rhomboid and surrounds a single cell. 

In all cases, the JP vesicle tank-treads in TG flow as it does
in the shear flow. The
direction of the tank-treading, however, varies with the cell size
(Figure~\ref{fig:BTG_Scale}(a)--(d), directed arcs). We attribute this
change in rotational orientation to the total rotational moment of the
cells that the vesicle passes through.

We calculate the vesicle area $A$ and length $L$ (initial area $A_0$,
initial length $L_0$) by averaging the area and length of the inner and
outer leaflets. As shown in Figure~\ref{fig:BTG_Scale}(e), the relative
area $A/A_0$ decreases during the simulations, with the rate depending
nonmonotonically on $l$. The relative length $L/L_0$, however, is
constant in $t$ for all cell sizes (Figure~\ref{fig:BTG_Scale}(e),
inset). This implies that the vesicle behaves as an inextensible,
permeable membrane. In previous work, \citet{Fu2022_JFM} determined the
permeability constant of particle-based vesicles in the context of shear
background flow~\cite{chabanon2017, qua-gan-you2021}.

\begin{figure}
  \begin{center}
\includegraphics[width=1.0\textwidth]{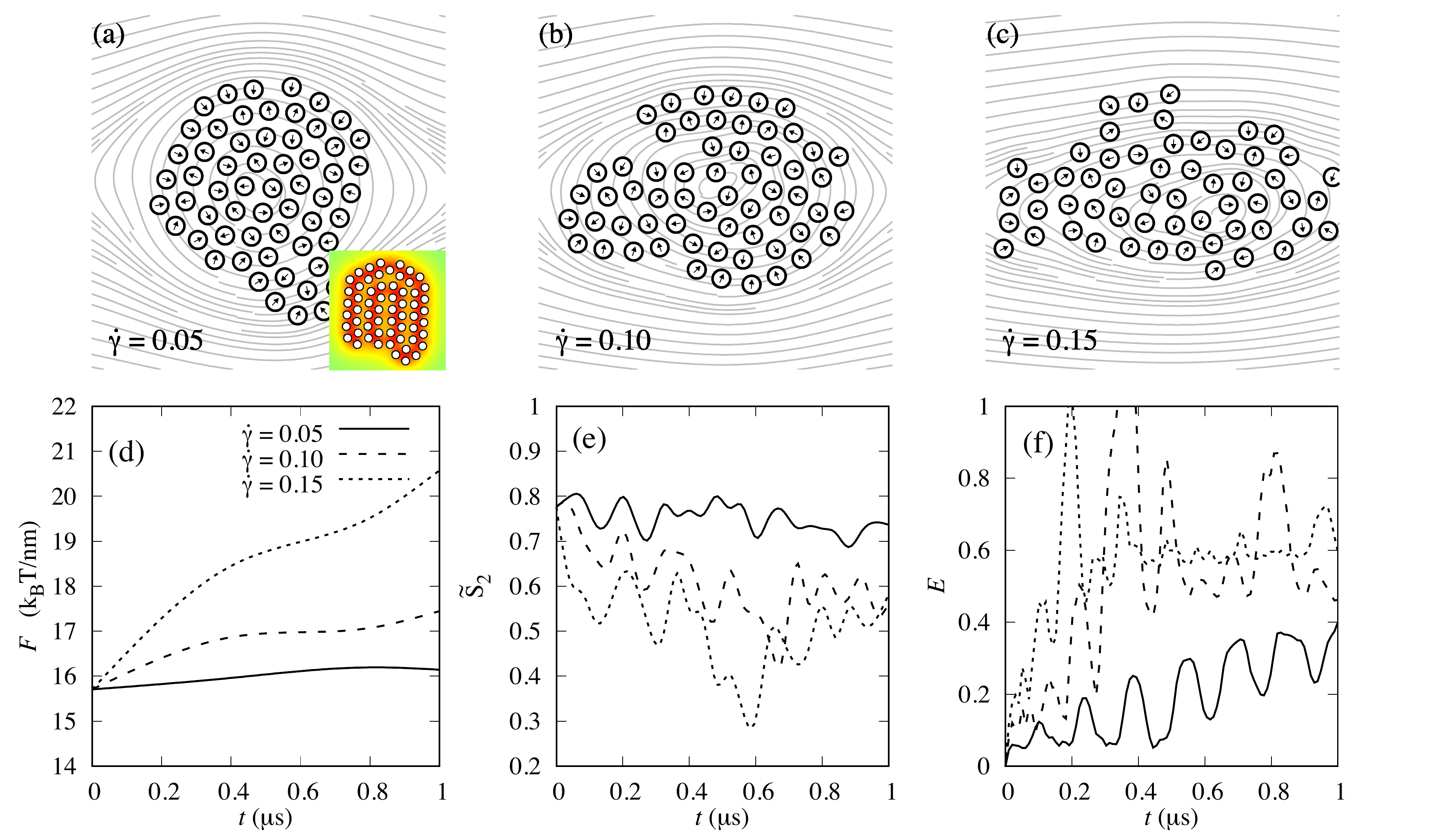}
  \end{center}
  \vspace{-20pt}  
  \caption{\label{fig:BC2_shear} A multilamellar structure in a shear
  flow. Panels (a)--(c) are snapshots for $\dot \gamma = \{0.05, 0.1,
  0.15\}$ at $t=0.6\ \upmu$s where the pre-relaxed initial configuration
  is shown in inset of panel (a). The streamlines are plotted in the
  background. Panel (d) shows the free energies; panel (e) shows
  orientational parameter $\tilde{S}_2$; panel (f) shows positional
  parameter E.}
\end{figure}

\subsection{Multilamellar and striated configurations in shear and TG flow}
\begin{figure}
  \begin{center}
\includegraphics[width=1.0\textwidth]{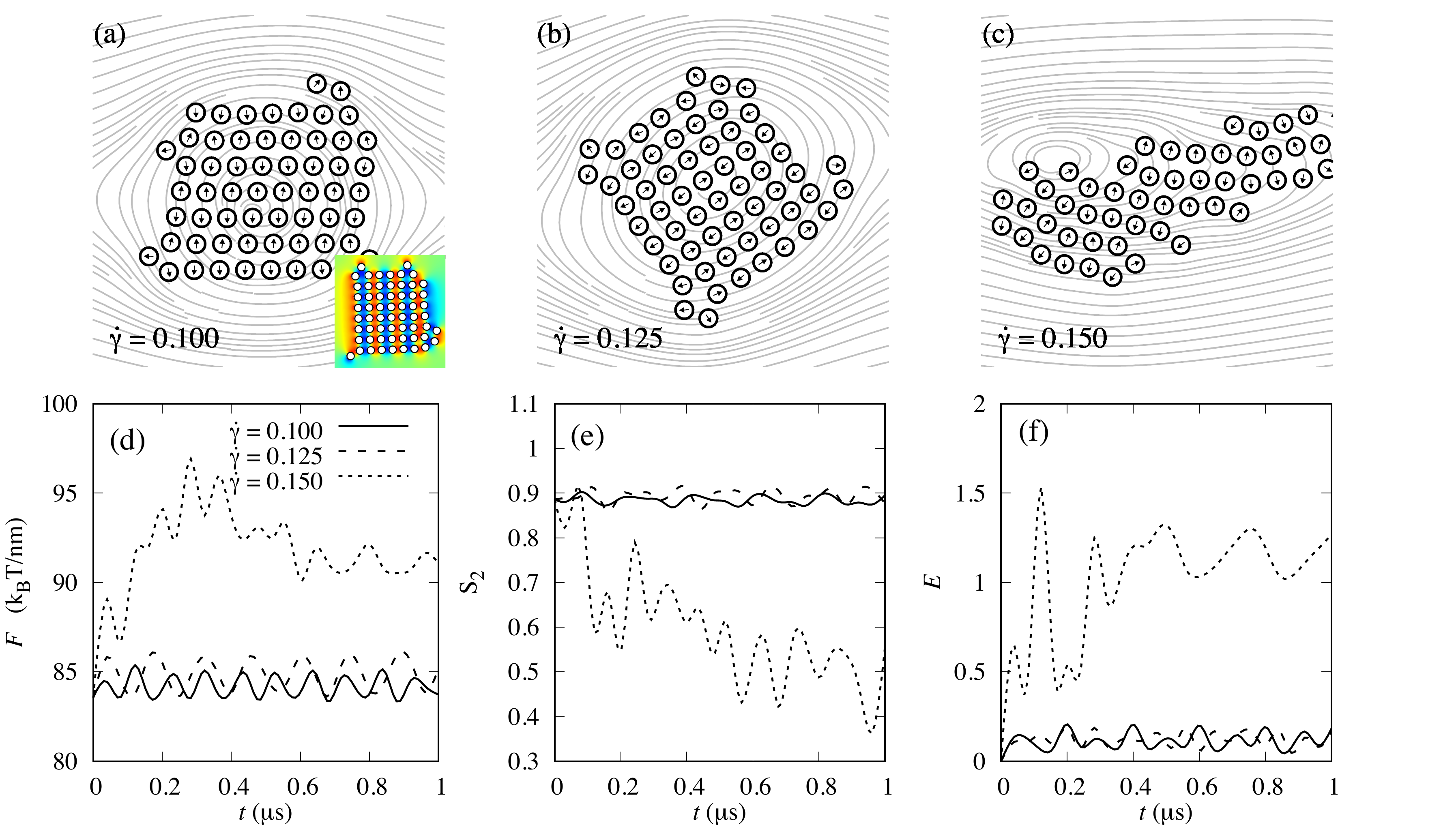}        
  \end{center}
  \caption{\label{fig:BC3_shear} A striated configuration in a shear
  flow. Panels (a)-(c) are snapshots for $\dot \gamma = \{0.1, 0.125,
  0.15\}$ at $t=0.1\mu$s where the pre-relaxed initial configuration is
  shown in inset of panel (a). The streamlines are plotted in the
  background. Panel (d) shows the free energies; panel (e) shows
  orientational parameter $\tilde{S}_2$; panel (f) shows positional
  parameter E.}
\end{figure}

We have simulated particles for the multilamellar (BC (ii)) and striated
(BC (iii)) boundary conditions, in both shear and TG background flow.
The results are qualitatively similar and will be presented in tandem.
For the multilamellar (BC (ii)) case, we place a 60-particle configuration (inset
of Figure~\ref{fig:BC2_shear}(a)) in shear flow with $\dot\gamma=0.05,
0.1, 0.15$ and in TG flow with $V_0=0.1, 0.15, 0.2$. For the
striated  (BC (iii)) case, we also place a 60-particle configuration (inset of
Figure~\ref{fig:BC3_shear}(a)) in shear and TG background flows. Here,
somewhat larger rates are required to see appreciable deformations for
this boundary condition: $\dot\gamma=0.1, 0.125, 0.15$ and $V_0=0.2,
0.25, 0.3$, respectively.
The BC (ii) and BC (iii) in shear flow cases are visualized in
the latter half of 
Supplementary Movie S2 (low shear rate) and 
Supplementary Movie S4 (high shear rate), respectively.
The low and high flow rates for TG flow are shown in
the latter half of
Supplementary Movie S3 and 
Supplementary Movie S5, respectively.

In terms of shear flow, the free energies are steady at the lowest shear
rates (Figure~\ref{fig:BC2_shear}(d), Figure~\ref{fig:BC3_shear}(d)).
The free energy of the striated configuration oscillates by $\pm 1$
$\mathrm{k_BT}$/nm due to a square reference region deforming into a
rhomboidal shape under shear flow (Figure~\ref{fig:BC3_shear}(d), solid
curve). No such oscillation is present for the multilamellar
configuration since this shape is circularly isotropic
(Figure~\ref{fig:BC2_shear}(d), solid curve). At the lowest shear rates,
there is, however, oscillation in the strains of both configurations,
while the order parameter $\tilde S_2$ is nearly constant
(Figure~\ref{fig:BC2_shear}(e),(f), Figure~\ref{fig:BC3_shear}(e),(f),
solid curves). In summary, both multilamellar and striated
configurations behave as nearly rigid bodies under shear flow when the
shear rate $\dot \gamma$ is low.

In the high shear rate cases ($\dot\gamma=0.15$), both the multilamellar (BC (ii))
and striated (BC (iii)) configurations become disordered. For BC (ii), the lamella
break apart so that individual bilayers are no longer discernible
(Figure~\ref{fig:BC2_shear}(c)). For BC (iii), the stria peel away from
the main body, but remain individually intact
(Figure~\ref{fig:BC3_shear}(c)). Overall, the particles in BC (ii) and
BC (iii) remain bounded and do not drift away in the shear background
flow like for the BC (i) (Figure~\ref{fig:BC1_shear}(c),
Figure~\ref{fig:Ves_shear}(c)).

\begin{figure}
  \begin{center}
\includegraphics[width=1.0\textwidth]{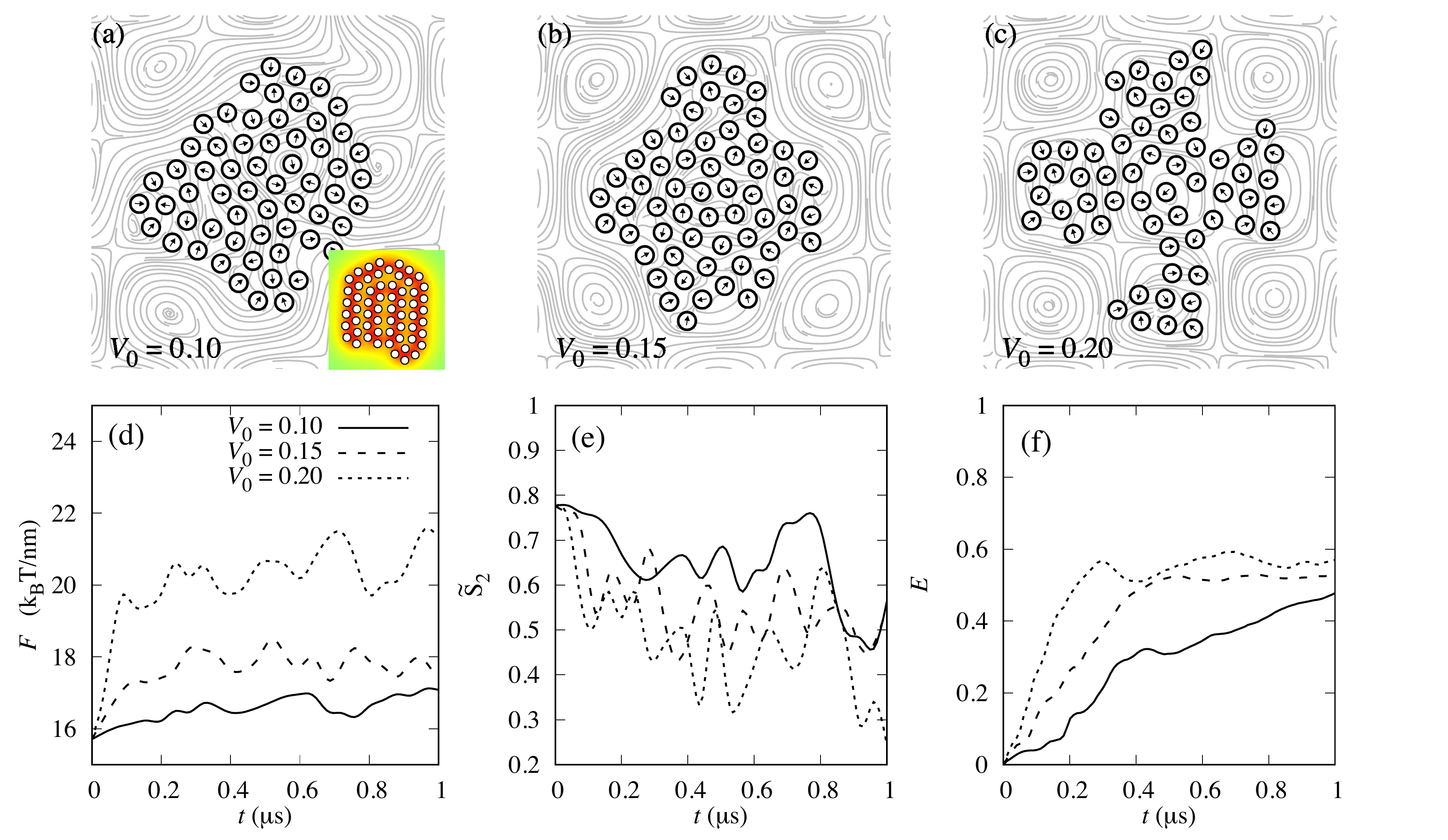} 
  \end{center}
  \vspace{-20pt}
  \caption{\label{fig:BC2_TG} A multilamellar structure in a
  Taylor-Green flow. Panels (a)-(c) are snapshots for $V_0 = \{0.1,
  0.15, 0.2\}$ at $t=0.6\ \mu$s where the pre-relaxed initial
  configuration is shown in inset of panel (a). The streamlines are
  plotted in the background. Panel (d) shows the free energies; panel
  (e) shows orientational parameter $\tilde{S}_2$; panel (f) shows
  positional parameter E.}
\end{figure}

\begin{figure}
  \begin{center}
\includegraphics[width=1.0\textwidth]{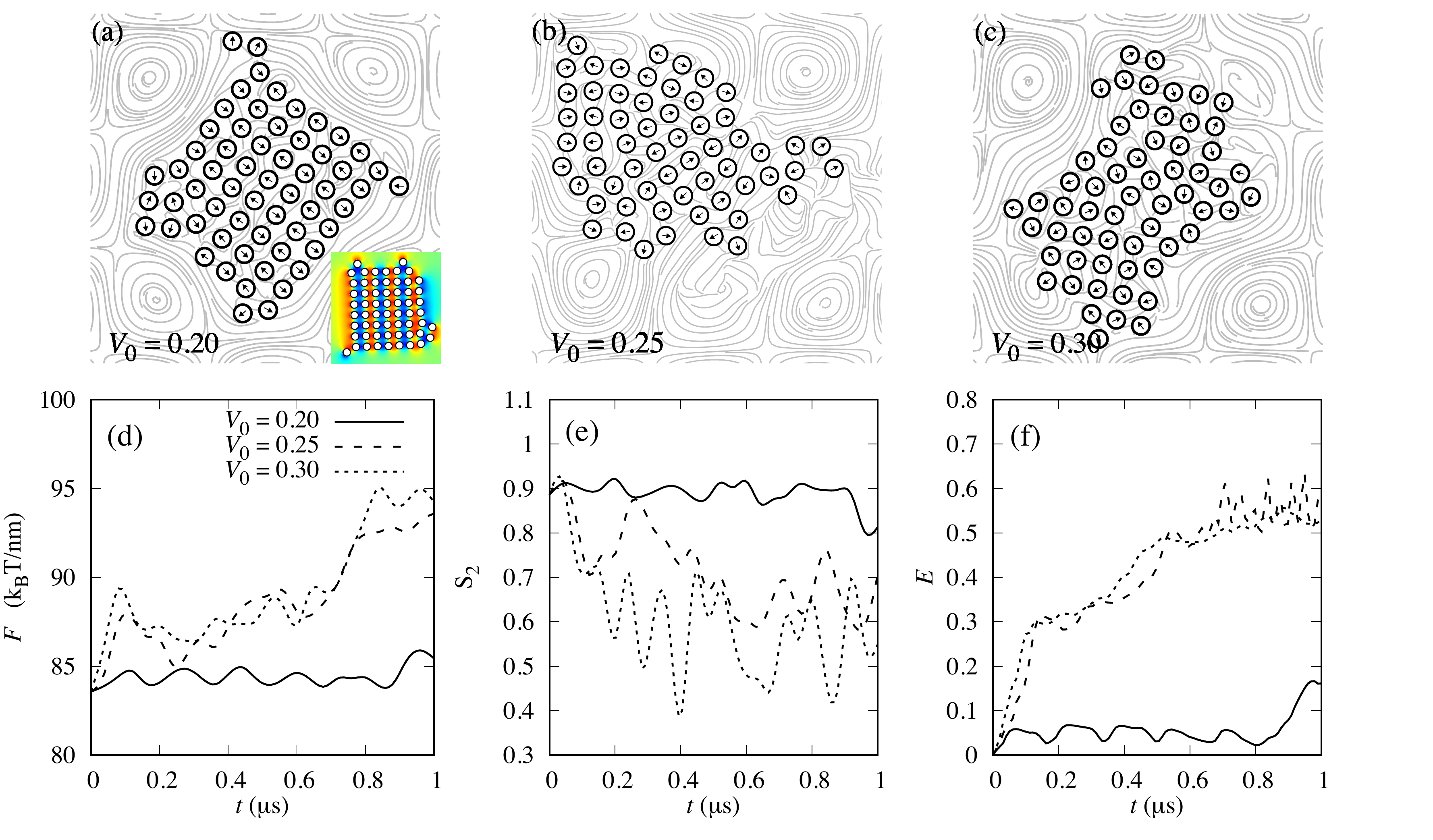}            
  \end{center}
  \vspace{-20pt}  
  \caption{\label{fig:BC3_TG} A striated configuration in a Taylor-Green
  flow. Panels (a)-(c) are snapshots for $V_0=\{0.2, 0.25, 0.3\}$ at
  $t=0.4\mu$s where the pre-relaxed initial configuration is shown in
  inset of panel (a). The streamlines are plotted in the background.
  Panel (d) shows the free energies; panel (e) shows orientational
  parameter $\tilde{S}_2$; panel (f) shows positional parameter E.}
\end{figure}

Under a TG background flow, the free energy $F$ is also basically
constant for the lowest flow rates $V_0$ (Figure~\ref{fig:BC2_TG}(d),
Figure~\ref{fig:BC3_TG}(d), solid curves). The striated configuration is
unperturbed by the flow, and is engulfed by a single, larger, rotating
flow cell (Figure~\ref{fig:BC3_TG}(a)). The multilamellar configuration
is perturbed by the flow, as seen in the increase in the strain
parameter $E$ (Figure~\ref{fig:BC3_TG}(f), solid curve). The difference
in response to the background flow suggests that the multilamellar
configuration allows for local rearrangement of the particles while
retaining the overall shape.

At large flow rates $V_0$, both configurations depart significantly from
their local equilibrium. The lamella bilayers from BC (ii) break apart,
forming several unlayered bilayer components
(Figure~\ref{fig:BC2_TG}(c)). In BC (iii), the stria also break apart,
but neighboring particles form `X'-like arrangements
(Figure~\ref{fig:BC3_TG}(d)), resembling the doubly alternating director
equilibrium (Figure~\ref{fig:relax}(d), center, top white rectangle).


\section{Discussion\label{sec:discussion}}
Results in \S~\ref{sec:results} are from simulations with $N_b = 60$ particles.
This particle number was
large enough for the JP to form measurable structures while giving
manageable simulation times (days) on a single computing processor. The
simulation time for $N_b = 198$ in Figure~\ref{fig:relax} and
Figure~\ref{fig:relax_energy} was several weeks, and so we avoided
simulations of this size for our results. We used $N = 16$ grid points
per particle for all simulations. We performed a convergence study and
found that using $N = 16$ or $N = 24$ grid points per particle gave
quantitatively indistinguishable results. The time courses consisted of
$O(10^3)$ time steps with $\Delta t = 0.2$ ns.

The free energy $F$ defined in~\eqref{eq:free_energy} scales with the
boundary data $g$ and repulsion modulus $M$. 
We thus scale all boundary data so that the
square integral of $g$ is independent of the type of boundary condition; BC
(i), (ii), or (iii). This way, for a fixed configuration, the
hydrophobic interaction portion of the free energy (the integral
in~\eqref{eq:free_energy}) converges to $\gamma N_b$ in the limit
$\rho = 0$. This gives a free energy per particle that is independent of
the shape and intensity of the hydrophobic interface. In practice, $\rho
> 0$ is fixed and the JPs assume different configurations. As a result,
the equilibrium energies per particle in our simulations are different
and depend on the boundary condition (Figure~\ref{fig:relax_energy}).
They are about 0.26 $\mathrm{k_BT}$/nm for BC (ii), 0.55
$\mathrm{k_BT}$/nm for BC (i), and 1.4 $\mathrm{k_BT}$/nm for BC (iii).

Next, we impose a flow on the JP suspension to study
deformations of JP assemblies as a function of flow strength. There are three categories
of boundary conditions---BC (i), BC (ii), and BC (iii)---that form
bilayer, multilamellar, and striated configurations, respectively. For
BC (i), we consider both an unstructured bilayer and a vesicle
bilayer. The background flows are a linear shear flow and a steady TG flow.

We characterize the effective material properties of the JP
structures based on the changes in free energy, orientational order, and
strain as a function of flow strength. Generally, $F$ increases, $\tilde
S_2$ decreases, and $E$ increases with increases in both $\dot \gamma$
and $V_0$, as expected. Structurally, the striated JP configuration for
BC (iii) behaves the most like a rigid body over the flow strengths
tested. For this structure, the deformation measures are constant in
time for shear rates $\dot \gamma$ up to $0.15$ and TG flow rate $V_0$
up to $0.25$ (Figure~\ref{fig:BC3_shear} and Figure~\ref{fig:BC3_TG}).
In contrast, a $25\%$--$50 \%$ smaller value for $\dot \gamma$ or $V_0$ is
required to observe nonrigid deformations for the BC (i) and BC (ii)
cases.

In terms of fluid behavior, the multilamellar (BC (ii)) structure
behaves as a shear thinning fluid while the striated structure (BC (iii))
has a finite yield stress. Qualitatively, the strains for BC (ii)
increases gradually with shear rate (Figure~\ref{fig:BC2_shear}(f)). In
contrast, the strains for BC (iii) are constant over a range of flow
strengths and then increase when the flow strengths are large enough to
deform the body (Figure~\ref{fig:BC3_shear}(f)). To quantify this
behavior, we form 
\begin{align}
\label{eq:mu_asm}
\mu_{\text{asm}} = \frac{\dot \gamma}{\dot E} \mu 
\end{align}
as an effective dynamic viscosity of the particle assemblies. The
numerator $\dot \gamma \mu$ gives the force per area provided by the
solvent stresses where $\mu = 1$ mPa s is the solvent viscosity. The
strain rate $\dot E$ of the body is the slope of a linear fit to the
strain curves e.g., in Figure~\ref{fig:BC2_shear}(f).

The viscosities of particle assemblies calculated from \eqref{eq:mu_asm} are shown in
Table~\ref{tbl:bcii_visc}. The middle row gives the solid area fraction
$\phi$; the sum of the rigid particle areas divided by the total area of
the solvent and particle region. We observe that for the multilamellar
structure, $\mu_{\text{asm}}$ decreases with increasing shear rate. The
predicted viscosities of the particle assemblies are on the order of
hundreds of times that of water, consistent with suspension viscosities
for noninteracting particles~\cite{KONIJN201461}. In the case of
striated structures, viscosity is effectively infinite for the lower two
of the shear rates, and then drops to a finite value $\mu_{\text{asm}} =
60$ mPa s when the shear rate reaches $0.15$ ns$^{-1}$.
\begin{table}
  \caption{\label{tbl:bcii_visc} The dynamic viscosity of assemblies
  $\mu_{\text{asm}}$ (mPa s).}
\centering
\begin{tabularx}{0.7\textwidth}{c|X|X|X||X|X|X}
&\multicolumn{3}{c||}{BC (ii)} & \multicolumn{3}{c}{BC (iii)}\\
\hline
  $\dot \gamma$ & 0.05 & 0.10 \quad & 0.15 & 0.100 & 0.125 & 0.15\\
  \hline
  $\phi$ & 0.50 & 0.47 & 0.41 & 0.47 & 0.46 & 0.45 \\
  \hline
  $\mu_{\text{sus}} $ & 401 & 66 & 42 & 2.5e3 & 1.4e3 & 60\\
\hline
\end{tabularx}
\end{table}

In the case of BC (i), our previous work~\cite{Fu2022_JFM} calculated a
friction coefficient $b$ for the intermonolayer slip between the two
leaflets of a bilayer. \citet{denOtter2007,Zgorski2019} and
\citet{doi:10.1073/pnas.2100156118} have calculated in-plane viscosities
for fully three-dimensional bilayers. Although not part of this work, it
is in principle possible to obtain an in-plane viscosity using the
hydrophobic potential~\eqref{eq:free_energy} by replacing the boundary
condition~\eqref{eq:bc-type} with one that is everywhere constant, in
effect modeling lipids in a bilayer as an array of elongated, purely
hydrophobic pillars. 

While increased flow strengths generally injected energy into the
suspension, the simulation results show that the unstructured bilayer
actually increases orientational order and has somewhat lower free
energy under low shear rates (Figure~\ref{fig:BC1_shear}(d,f)). This
effect suggests that in amphiphilic suspensions, it might be possible to
decrease the excess area of the system by subjecting the suspension to a
moderate shear flow (Supplementary Movie S2, right panel).
Not all background flows produce this
``organizing'' effect, as it is not observed in the TG background flow
case (Figure~\ref{fig:BC1_TG}, Supplementary Movie S3, right panel).
\citet{PhysRevLett.128.256102} have
shown how to drive particle clusters into arbitrary target shapes by
solving a first passage problem.

The characteristics of the self-assembly of amphiphilic JP into
onion-like dendrimersomes like for BC (ii) were previously studied by molecular dynamics
simulations~\cite{C9NR05885K}. The molecular dynamics simulations use an
anisotropic pair potential to describe the particle interactions
where a harmonic form describes the repulsive part and an anisotropic
attractive part describes the hydrophobic interactions~\cite{HongCacciutoLuijtenGranick2008}.
We also use a pair potential $P$ in the free energy \eqref{eq:free_energy} for repulsion,
whereas the long-range, hydrophobic attraction \eqref{eq:FandTdef}
is nonadditive and comes from domain-dependent variation of
the water order parameter $u(\xx)$ in the bulk.

\citet{kohl-cor-che-vee22} have simulated JP suspensions in three
dimensions using the same free energy formulation of hydrophobic
attraction used in the present work. They observed spontaneous
aggregation into micelles generically and formation of bilayers when the
initial configuration was chosen sufficiently close to the final
configuration. They also studied self-assembly of bipolar electric JP,
which corresponds to our BC (iii) JP (but electric potential, instead of a water order parameter),
to define the interaction.  Due to the differing interaction,
bipolar electric particles form parallel chains that
repel. In contrast, the hydrophobic interaction between stria is
attractive. The existence of 'X'-shaped local arrangements as a local
equilibrium (Figure~\ref{fig:relax}(d), top, middle rectangle) further
suggests there may be greater diversity in the set of possible
equilibria in three-dimensions when employing hydrophobic attraction.

\section{Conclusion}
\label{sec:conclusion}

In this work, we employ the newly developed JP model using
BIEs~\cite{Fu20, Fu2022_JFM} and tune the boundary conditions with
energy normalization to study the collective dynamics of amphiphilic
(BC(i)), biased hydrophobic (BC (ii)), and bipolar (BC(iii)) JPs under
various flowing conditions (quiescent flow, linear shear flow, and
Taylor-Green flow). Three quantities are computed to characterize the
dynamics of the collective configurations of JPs: free energy $F$,
strain parameter $E$, and scalar order parameter $S_2$. 
Under a given flow, we use these three measures of deformations to
quantify the differences in the collective dynamics between the three
types of JPs. These results, summarized below, provide general insight
into the dynamic control of active particles in a viscous suspension.

Within the framework of boundary integral equations, we derive a
convenient expression for computing the free energy using the single
layer potential (see equation~\eqref{eq:normal_deriv}). In a quiescent
flow, the free energy profiles demonstrate that the relaxation process
for particles confined in a certain size of box is independent of the
number of particles. However, we find that the final configurations does
depend on the initial distribution of particle directors $\dd_i$.
Therefore, multiple patterns or local energy minimum states may appear
depending on the initial setup.

Under a relatively weak linear shear flow, the amphiphilic JPs behave as
a unilamellar vesicle that elongates and tank-treads, with the scalar
order parameter increasing over time. The assembly of multi-lamella and
striated JP structures resemble a rigid body motion with minimal
deformation. The effective viscosity of the material quantitatively
validates this result. High shear-rate cases provide a range of critical
shear rates where the structures break apart and undergo topological
changes. We further show that free energy, scalar order parameter, and
strain are effective measures to quantitatively capture the collective hydrodynamics of
JPs under a linear shear flow.

Under a Taylor-Green flow, the amphiphilic JPs are the most interesting,
exhibiting different vesicle shapes depending on the ratio of the size
of the TG flow to the vesicle size. These results show that the shape of
the vesicle, whether square or polygonal, can be controlled by adjusting
the size of the cell in the TG flow. On the other hand, the assembly of
multi-lamella JPs and the striated JPs behave more like a rigid body
with connected subdomains, and the number of subdomains increases with
increasing strength of the TG flow. Overall, the multilamellar (BC (ii))
JP assembly behaves as a shear-thinning fluid, while the striated (BC
(iii)) JP assembly  possesses a yield stress.

The results reported here provide an inroad modeling framework for
hydrodynamics of active colloids~\cite{Meredithetal2022,
McGlassonBradley2021, Vutukuri2020, Mallory2017}. The present study also
helps us understand the rheology of JP oligomers that may be realized in
the experiments. We are extending this study to three dimensional
systems with more realistic features such as size distributions of JPs
and thermal fluctuations~\cite{kohl-cor-che-vee22}. From a numerical
perspective, it is straightforward to include random perturbations in
the particle shape and boundary condition that mimic interfacial
properties found in lab conditions~\cite{Bradley2016, Bradley2017,
Zarzaretal2015, doi:10.1021/la503455h}.

\section{Appendix}
\label{sec:appendix}
Our calculation of the free energy $F$ relied on converting the body
integral in~\eqref{eq:free_energy} into the surface integral
in~\eqref{eq:free_energy2} involving only single layer potentials and
tangential derivatives. This section supplies a proof of
identity~\eqref{eq:normal_deriv} used in this conversion.  Let 
\begin{equation}
  \label{eq:SLP}
  \mathcal{S}[\sigma](\xx) = \int_\Gamma G(\xx-\yy) \sigma(\yy)\, \dif s_\yy
\end{equation}
be the single layer potential for a density function $\sigma.$ Fix $\xx
\in \Gamma$, let $\nnu_{\xx} = \nnu(\xx)$ and $\nnu_{\yy}$ be the unit
normal at $\xx$, respectively $\yy$, in $\Gamma$, and let $\zz \in
\Omega$. The subscripts in $\nabla_{\zz}$ and $\nabla_{\yy}$ denote
differentiation with respect to $\zz$, respectively $\yy$.

Recall from~\eqref{eq:SL_BIE} that $u = \mathcal{D}[\sigma]$.
Then  
\begin{align*}
\nabla_{\zz} u(\zz) \cdot \nnu_{\xx}
&=\nnu_\xx \cdot \nabla_\zz \int_\Gamma \frac{\partial G(\zz-\yy)}{\partial \nnu_\yy}\sigma(\yy) \,\dif s_\yy\\
&=\int_\Gamma \nnu_\xx^\top \left(\nabla_\zz\nabla_\yy^\top  G(\zz-\yy)\right) \nnu_\yy\sigma(\yy)  \,\dif s_\yy\\
  &=-\int_\Gamma \nnu_\xx^\top \left(\nabla_\yy\nabla_\yy^\top G(\zz-\yy)\right)\nnu_\yy\sigma(\yy)\ \dif s_\yy,
\end{align*}
since we can interchange $\nabla_\zz$ with $-\nabla_\yy$.
Following~\citet{Hsiao2008}, \S 1.2,
\begin{align}
\nnu_\xx^\top \left(\nabla_\yy\nabla_\yy^\top G(\zz-\yy)\right)\nnu_\yy
=
-\tt_{\xx}^\top \left(\nabla_\yy\nabla_\yy^\top G(\zz-\yy)\right)\tt_{\yy}
+ \Delta_{\yy}G(\zz-\yy) \tt_{\xx}\cdot \tt_{\yy}.
\end{align}
Then, using that $\Delta_{\yy} G(\zz-\yy) = \rho^{-2} G(\zz-\yy)$,
interchanging $\nabla_\yy$ with $-\nabla_\zz$ once more, and integrating
by parts in arclength $s$, we obtain 
\begin{align*}
\nabla_{\zz} u(\zz) \cdot \nnu_{\xx}
&=-\int_\Gamma  \Delta_\yy G(\zz-\yy) {\bf t}_\xx \cdot   {\bf t}_\yy \sigma(\yy)\ \dif s_\yy
+\int_\Gamma({\bf t}_\xx\cdot\nabla_\yy)({\bf t}_\yy\cdot\nabla_\yy G(\zz-\yy))\sigma(\yy)\ \dif s_\yy\\
&= -\int_\Gamma \frac{1}{\rho^{2}} G(\zz-\yy) {\bf t}_\xx \cdot {\bf t}_\yy \sigma(\yy)\ \dif s_\yy  
-({\bf t}_\xx\cdot\nabla_\zz)\int_\Gamma \frac{\dif}{\dif s_\yy}G(\zz-\yy) \sigma(\yy)\ \dif s_\yy\\
&= -\frac{1}{\rho^2} {\bf t}_\xx\cdot \int_\Gamma G(\zz-\yy){\bf t}_\yy \sigma(\yy)\ \dif s_\yy + 
({\bf t}_\xx \cdot \nabla_\zz)\int_\Gamma G(\zz-\yy)\frac{\dif }{\dif s} \sigma(\yy)  \dif s_\yy.
\end{align*}
Letting $\zz\to\xx\in\Gamma$, and noting that both sides of the equation
are continuous, we obtain~\eqref{eq:normal_deriv}.

\begin{acknowledgments}
We thank useful conversations with E. Corona, M. Rachh and S. Jiang.
B.Q.~acknowledges support from the Simons Foundation, Mathematics and Physical Sciences-Collaboration Grants for Mathematicians, Award No. 527139.
Y.-N.Y.~acknowledges support from NSF (Grants No. DMS 1614863 and No. DMS 195160) and Flatiron Institute, part of Simons Foundation.

\end{acknowledgments}

\providecommand{\noopsort}[1]{}\providecommand{\singleletter}[1]{#1}%
%


\begin{thebibliography}{58}%
\makeatletter
\providecommand \@ifxundefined [1]{%
 \@ifx{#1\undefined}
}%
\providecommand \@ifnum [1]{%
 \ifnum #1\expandafter \@firstoftwo
 \else \expandafter \@secondoftwo
 \fi
}%
\providecommand \@ifx [1]{%
 \ifx #1\expandafter \@firstoftwo
 \else \expandafter \@secondoftwo
 \fi
}%
\providecommand \natexlab [1]{#1}%
\providecommand \enquote  [1]{``#1''}%
\providecommand \bibnamefont  [1]{#1}%
\providecommand \bibfnamefont [1]{#1}%
\providecommand \citenamefont [1]{#1}%
\providecommand \href@noop [0]{\@secondoftwo}%
\providecommand \href [0]{\begingroup \@sanitize@url \@href}%
\providecommand \@href[1]{\@@startlink{#1}\@@href}%
\providecommand \@@href[1]{\endgroup#1\@@endlink}%
\providecommand \@sanitize@url [0]{\catcode `\\12\catcode `\$12\catcode
  `\&12\catcode `\#12\catcode `\^12\catcode `\_12\catcode `\%12\relax}%
\providecommand \@@startlink[1]{}%
\providecommand \@@endlink[0]{}%
\providecommand \url  [0]{\begingroup\@sanitize@url \@url }%
\providecommand \@url [1]{\endgroup\@href {#1}{\urlprefix }}%
\providecommand \urlprefix  [0]{URL }%
\providecommand \Eprint [0]{\href }%
\providecommand \doibase [0]{https://doi.org/}%
\providecommand \selectlanguage [0]{\@gobble}%
\providecommand \bibinfo  [0]{\@secondoftwo}%
\providecommand \bibfield  [0]{\@secondoftwo}%
\providecommand \translation [1]{[#1]}%
\providecommand \BibitemOpen [0]{}%
\providecommand \bibitemStop [0]{}%
\providecommand \bibitemNoStop [0]{.\EOS\space}%
\providecommand \EOS [0]{\spacefactor3000\relax}%
\providecommand \BibitemShut  [1]{\csname bibitem#1\endcsname}%
\let\auto@bib@innerbib\@empty
\bibitem [{\citenamefont {Kirillova}\ \emph {et~al.}(2019)\citenamefont
  {Kirillova}, \citenamefont {Marschelke},\ and\ \citenamefont
  {Synytska}}]{KirillovaMarschelkeSynytska2019}%
  \BibitemOpen
  \bibfield  {author} {\bibinfo {author} {\bibfnamefont {A.}~\bibnamefont
  {Kirillova}}, \bibinfo {author} {\bibfnamefont {C.}~\bibnamefont
  {Marschelke}},\ and\ \bibinfo {author} {\bibfnamefont {A.}~\bibnamefont
  {Synytska}},\ }\href@noop {} {\bibfield  {journal} {\bibinfo  {journal} {ACS
  Applied Materials \& Interfaces}\ }\textbf {\bibinfo {volume} {11}},\
  \bibinfo {pages} {9643} (\bibinfo {year} {2019})}\BibitemShut {NoStop}%
\bibitem [{\citenamefont {Meredith}\ \emph {et~al.}(2022)\citenamefont
  {Meredith}, \citenamefont {Castonguay}, \citenamefont {Chiu}, \citenamefont
  {Brooks}, \citenamefont {Moerman}, \citenamefont {Torab}, \citenamefont
  {Wong}, \citenamefont {Sen}, \citenamefont {Velegol},\ and\ \citenamefont
  {Zarzar}}]{Meredithetal2022}%
  \BibitemOpen
  \bibfield  {author} {\bibinfo {author} {\bibfnamefont {C.~H.}\ \bibnamefont
  {Meredith}}, \bibinfo {author} {\bibfnamefont {A.~C.}\ \bibnamefont
  {Castonguay}}, \bibinfo {author} {\bibfnamefont {Y.-J.}\ \bibnamefont
  {Chiu}}, \bibinfo {author} {\bibfnamefont {A.~M.}\ \bibnamefont {Brooks}},
  \bibinfo {author} {\bibfnamefont {P.~G.}\ \bibnamefont {Moerman}}, \bibinfo
  {author} {\bibfnamefont {P.}~\bibnamefont {Torab}}, \bibinfo {author}
  {\bibfnamefont {P.~K.}\ \bibnamefont {Wong}}, \bibinfo {author}
  {\bibfnamefont {A.}~\bibnamefont {Sen}}, \bibinfo {author} {\bibfnamefont
  {D.}~\bibnamefont {Velegol}},\ and\ \bibinfo {author} {\bibfnamefont {L.~D.}\
  \bibnamefont {Zarzar}},\ }\href@noop {} {\bibfield  {journal} {\bibinfo
  {journal} {Matter}\ }\textbf {\bibinfo {volume} {5}},\ \bibinfo {pages} {616}
  (\bibinfo {year} {2022})}\BibitemShut {NoStop}%
\bibitem [{\citenamefont {Bradley}\ \emph {et~al.}(2017)\citenamefont
  {Bradley}, \citenamefont {Chen}, \citenamefont {Stebe},\ and\ \citenamefont
  {Lee}}]{Bradley2017}%
  \BibitemOpen
  \bibfield  {author} {\bibinfo {author} {\bibfnamefont {L.~C.}\ \bibnamefont
  {Bradley}}, \bibinfo {author} {\bibfnamefont {W.-H.}\ \bibnamefont {Chen}},
  \bibinfo {author} {\bibfnamefont {K.~J.}\ \bibnamefont {Stebe}},\ and\
  \bibinfo {author} {\bibfnamefont {D.}~\bibnamefont {Lee}},\ }\href@noop {}
  {\bibfield  {journal} {\bibinfo  {journal} {Current Opinion in Colloid \&
  Interface Science}\ }\textbf {\bibinfo {volume} {30}},\ \bibinfo {pages} {25}
  (\bibinfo {year} {2017})}\BibitemShut {NoStop}%
\bibitem [{\citenamefont {Kang}\ and\ \citenamefont
  {Honciuc}(2018)}]{KangHonciuc2018}%
  \BibitemOpen
  \bibfield  {author} {\bibinfo {author} {\bibfnamefont {C.}~\bibnamefont
  {Kang}}\ and\ \bibinfo {author} {\bibfnamefont {A.}~\bibnamefont {Honciuc}},\
  }\href@noop {} {\bibfield  {journal} {\bibinfo  {journal} {The Journal of
  Physical Chemistry Letters}\ }\textbf {\bibinfo {volume} {9}},\ \bibinfo
  {pages} {1415} (\bibinfo {year} {2018})}\BibitemShut {NoStop}%
\bibitem [{\citenamefont {Hong}\ \emph {et~al.}(2008)\citenamefont {Hong},
  \citenamefont {Cacciuto}, \citenamefont {Luijten},\ and\ \citenamefont
  {Granick}}]{HongCacciutoLuijtenGranick2008}%
  \BibitemOpen
  \bibfield  {author} {\bibinfo {author} {\bibfnamefont {L.}~\bibnamefont
  {Hong}}, \bibinfo {author} {\bibfnamefont {A.}~\bibnamefont {Cacciuto}},
  \bibinfo {author} {\bibfnamefont {E.}~\bibnamefont {Luijten}},\ and\ \bibinfo
  {author} {\bibfnamefont {S.}~\bibnamefont {Granick}},\ }\href@noop {}
  {\bibfield  {journal} {\bibinfo  {journal} {Langmuir}\ }\textbf {\bibinfo
  {volume} {24}},\ \bibinfo {pages} {621} (\bibinfo {year} {2008})}\BibitemShut
  {NoStop}%
\bibitem [{\citenamefont {Gheisari}\ \emph {et~al.}(2021)\citenamefont
  {Gheisari}, \citenamefont {Shafiee}, \citenamefont {Abbasi}, \citenamefont
  {Jangjou}, \citenamefont {Izadpanah}, \citenamefont {Vaez},\ and\
  \citenamefont {Amani}}]{GheisariSahfieeAbbasiEtAl2021_DMR}%
  \BibitemOpen
  \bibfield  {author} {\bibinfo {author} {\bibfnamefont {F.}~\bibnamefont
  {Gheisari}}, \bibinfo {author} {\bibfnamefont {M.}~\bibnamefont {Shafiee}},
  \bibinfo {author} {\bibfnamefont {M.}~\bibnamefont {Abbasi}}, \bibinfo
  {author} {\bibfnamefont {A.}~\bibnamefont {Jangjou}}, \bibinfo {author}
  {\bibfnamefont {P.}~\bibnamefont {Izadpanah}}, \bibinfo {author}
  {\bibfnamefont {A.}~\bibnamefont {Vaez}},\ and\ \bibinfo {author}
  {\bibfnamefont {A.~M.}\ \bibnamefont {Amani}},\ }\href@noop {} {\bibfield
  {journal} {\bibinfo  {journal} {Drug Metabolism Reviews}\ }\textbf {\bibinfo
  {volume} {53}},\ \bibinfo {pages} {592} (\bibinfo {year} {2021})}\BibitemShut
  {NoStop}%
\bibitem [{\citenamefont {Liu}\ \emph {et~al.}(2016)\citenamefont {Liu},
  \citenamefont {Yang}, \citenamefont {Huang}, \citenamefont {Huang},
  \citenamefont {Zhang}, \citenamefont {Deng}, \citenamefont {Wang},
  \citenamefont {Zhou}, \citenamefont {Liu}, \citenamefont {Kalish},
  \citenamefont {Khachab}, \citenamefont {Chen},\ and\ \citenamefont
  {Nie}}]{LiuYangHuangEtAl2016_Angew}%
  \BibitemOpen
  \bibfield  {author} {\bibinfo {author} {\bibfnamefont {Y.}~\bibnamefont
  {Liu}}, \bibinfo {author} {\bibfnamefont {X.}~\bibnamefont {Yang}}, \bibinfo
  {author} {\bibfnamefont {Z.}~\bibnamefont {Huang}}, \bibinfo {author}
  {\bibfnamefont {P.}~\bibnamefont {Huang}}, \bibinfo {author} {\bibfnamefont
  {Y.}~\bibnamefont {Zhang}}, \bibinfo {author} {\bibfnamefont
  {L.}~\bibnamefont {Deng}}, \bibinfo {author} {\bibfnamefont {Z.}~\bibnamefont
  {Wang}}, \bibinfo {author} {\bibfnamefont {Z.}~\bibnamefont {Zhou}}, \bibinfo
  {author} {\bibfnamefont {Y.}~\bibnamefont {Liu}}, \bibinfo {author}
  {\bibfnamefont {H.}~\bibnamefont {Kalish}}, \bibinfo {author} {\bibfnamefont
  {N.~M.}\ \bibnamefont {Khachab}}, \bibinfo {author} {\bibfnamefont
  {X.}~\bibnamefont {Chen}},\ and\ \bibinfo {author} {\bibfnamefont
  {Z.}~\bibnamefont {Nie}},\ }\href@noop {} {\bibfield  {journal} {\bibinfo
  {journal} {Angew. Chem. Int. Ed}\ }\textbf {\bibinfo {volume} {55}},\
  \bibinfo {pages} {15297} (\bibinfo {year} {2016})}\BibitemShut {NoStop}%
\bibitem [{\citenamefont {Li}\ \emph {et~al.}(2019)\citenamefont {Li},
  \citenamefont {Wang}, \citenamefont {Yao}, \citenamefont {Yu}, \citenamefont
  {Yan},\ and\ \citenamefont {Zhang}}]{LiWangYaoEtAl2019_Nanoscale}%
  \BibitemOpen
  \bibfield  {author} {\bibinfo {author} {\bibfnamefont {J.}~\bibnamefont
  {Li}}, \bibinfo {author} {\bibfnamefont {J.}~\bibnamefont {Wang}}, \bibinfo
  {author} {\bibfnamefont {Q.}~\bibnamefont {Yao}}, \bibinfo {author}
  {\bibfnamefont {K.}~\bibnamefont {Yu}}, \bibinfo {author} {\bibfnamefont
  {Y.}~\bibnamefont {Yan}},\ and\ \bibinfo {author} {\bibfnamefont
  {J.}~\bibnamefont {Zhang}},\ }\href@noop {} {\bibfield  {journal} {\bibinfo
  {journal} {Nanoscale}\ }\textbf {\bibinfo {volume} {11}},\ \bibinfo {pages}
  {7221} (\bibinfo {year} {2019})}\BibitemShut {NoStop}%
\bibitem [{\citenamefont {Bradley}\ \emph {et~al.}(2016)\citenamefont
  {Bradley}, \citenamefont {Stebe},\ and\ \citenamefont {Lee}}]{Bradley2016}%
  \BibitemOpen
  \bibfield  {author} {\bibinfo {author} {\bibfnamefont {L.~C.}\ \bibnamefont
  {Bradley}}, \bibinfo {author} {\bibfnamefont {K.~J.}\ \bibnamefont {Stebe}},\
  and\ \bibinfo {author} {\bibfnamefont {D.}~\bibnamefont {Lee}},\ }\href@noop
  {} {\bibfield  {journal} {\bibinfo  {journal} {Journal of the American
  Chemical Society}\ }\textbf {\bibinfo {volume} {138}},\ \bibinfo {pages}
  {11437} (\bibinfo {year} {2016})}\BibitemShut {NoStop}%
\bibitem [{\citenamefont {Zarzar}\ \emph {et~al.}(2015)\citenamefont {Zarzar},
  \citenamefont {Sresht}, \citenamefont {Sletten}, \citenamefont {Kalow},
  \citenamefont {Blankschtein},\ and\ \citenamefont {Swager}}]{Zarzaretal2015}%
  \BibitemOpen
  \bibfield  {author} {\bibinfo {author} {\bibfnamefont {L.~D.}\ \bibnamefont
  {Zarzar}}, \bibinfo {author} {\bibfnamefont {V.}~\bibnamefont {Sresht}},
  \bibinfo {author} {\bibfnamefont {E.~M.}\ \bibnamefont {Sletten}}, \bibinfo
  {author} {\bibfnamefont {J.~A.}\ \bibnamefont {Kalow}}, \bibinfo {author}
  {\bibfnamefont {D.}~\bibnamefont {Blankschtein}},\ and\ \bibinfo {author}
  {\bibfnamefont {T.~M.}\ \bibnamefont {Swager}},\ }\href@noop {} {\bibfield
  {journal} {\bibinfo  {journal} {Nature}\ }\textbf {\bibinfo {volume} {518}},\
  \bibinfo {pages} {520} (\bibinfo {year} {2015})}\BibitemShut {NoStop}%
\bibitem [{\citenamefont {Hagan}\ and\ \citenamefont
  {Grason}(2021)}]{RevModPhys.93.025008}%
  \BibitemOpen
  \bibfield  {author} {\bibinfo {author} {\bibfnamefont {M.~F.}\ \bibnamefont
  {Hagan}}\ and\ \bibinfo {author} {\bibfnamefont {G.~M.}\ \bibnamefont
  {Grason}},\ }\href@noop {} {\bibfield  {journal} {\bibinfo  {journal} {Rev.
  Mod. Phys.}\ }\textbf {\bibinfo {volume} {93}},\ \bibinfo {pages} {025008}
  (\bibinfo {year} {2021})}\BibitemShut {NoStop}%
\bibitem [{\citenamefont {Collard}\ \emph {et~al.}(2020)\citenamefont
  {Collard}, \citenamefont {Grosjean},\ and\ \citenamefont
  {Vandewalle}}]{CollardGrosjeanVandewalle2020}%
  \BibitemOpen
  \bibfield  {author} {\bibinfo {author} {\bibfnamefont {Y.}~\bibnamefont
  {Collard}}, \bibinfo {author} {\bibfnamefont {G.}~\bibnamefont {Grosjean}},\
  and\ \bibinfo {author} {\bibfnamefont {N.}~\bibnamefont {Vandewalle}},\
  }\href@noop {} {\bibfield  {journal} {\bibinfo  {journal} {Communications
  Physics}\ }\textbf {\bibinfo {volume} {3}} (\bibinfo {year}
  {2020})}\BibitemShut {NoStop}%
\bibitem [{\citenamefont {Vutukuri}\ \emph {et~al.}(2020)\citenamefont
  {Vutukuri}, \citenamefont {Hoore}, \citenamefont {Abaurrea-Velasco},
  \citenamefont {van Buren}, \citenamefont {Dutto}, \citenamefont {Auth},
  \citenamefont {Fedosov}, \citenamefont {Gompper},\ and\ \citenamefont
  {Vermant}}]{Vutukuri2020}%
  \BibitemOpen
  \bibfield  {author} {\bibinfo {author} {\bibfnamefont {H.~R.}\ \bibnamefont
  {Vutukuri}}, \bibinfo {author} {\bibfnamefont {M.}~\bibnamefont {Hoore}},
  \bibinfo {author} {\bibfnamefont {C.}~\bibnamefont {Abaurrea-Velasco}},
  \bibinfo {author} {\bibfnamefont {L.}~\bibnamefont {van Buren}}, \bibinfo
  {author} {\bibfnamefont {A.}~\bibnamefont {Dutto}}, \bibinfo {author}
  {\bibfnamefont {T.}~\bibnamefont {Auth}}, \bibinfo {author} {\bibfnamefont
  {D.~A.}\ \bibnamefont {Fedosov}}, \bibinfo {author} {\bibfnamefont
  {G.}~\bibnamefont {Gompper}},\ and\ \bibinfo {author} {\bibfnamefont
  {J.}~\bibnamefont {Vermant}},\ }\href@noop {} {\bibfield  {journal} {\bibinfo
   {journal} {Nature}\ }\textbf {\bibinfo {volume} {586}},\ \bibinfo {pages}
  {52} (\bibinfo {year} {2020})}\BibitemShut {NoStop}%
\bibitem [{\citenamefont {Boccardo}\ and\ \citenamefont
  {Pierre-Louis}(2022)}]{PhysRevLett.128.256102}%
  \BibitemOpen
  \bibfield  {author} {\bibinfo {author} {\bibfnamefont {F.}~\bibnamefont
  {Boccardo}}\ and\ \bibinfo {author} {\bibfnamefont {O.}~\bibnamefont
  {Pierre-Louis}},\ }\href@noop {} {\bibfield  {journal} {\bibinfo  {journal}
  {Physical Review Letters}\ }\textbf {\bibinfo {volume} {128}},\ \bibinfo
  {pages} {256102} (\bibinfo {year} {2022})}\BibitemShut {NoStop}%
\bibitem [{\citenamefont {Manoharan}(2015)}]{Manoharan2015_Science}%
  \BibitemOpen
  \bibfield  {author} {\bibinfo {author} {\bibfnamefont {V.~N.}\ \bibnamefont
  {Manoharan}},\ }\href@noop {} {\bibfield  {journal} {\bibinfo  {journal}
  {Science}\ }\textbf {\bibinfo {volume} {349}},\ \bibinfo {pages} {1253751}
  (\bibinfo {year} {2015})}\BibitemShut {NoStop}%
\bibitem [{\citenamefont {Kirillova}\ \emph {et~al.}(2014)\citenamefont
  {Kirillova}, \citenamefont {Stoychev}, \citenamefont {Ionov},\ and\
  \citenamefont {Synytska}}]{doi:10.1021/la503455h}%
  \BibitemOpen
  \bibfield  {author} {\bibinfo {author} {\bibfnamefont {A.}~\bibnamefont
  {Kirillova}}, \bibinfo {author} {\bibfnamefont {G.}~\bibnamefont {Stoychev}},
  \bibinfo {author} {\bibfnamefont {L.}~\bibnamefont {Ionov}},\ and\ \bibinfo
  {author} {\bibfnamefont {A.}~\bibnamefont {Synytska}},\ }\href@noop {}
  {\bibfield  {journal} {\bibinfo  {journal} {Langmuir}\ }\textbf {\bibinfo
  {volume} {30}},\ \bibinfo {pages} {12765} (\bibinfo {year}
  {2014})}\BibitemShut {NoStop}%
\bibitem [{\citenamefont {Fu}\ \emph {et~al.}(2020)\citenamefont {Fu},
  \citenamefont {Ryham}, \citenamefont {Kl\"ockner}, \citenamefont {Wala},
  \citenamefont {Jiang},\ and\ \citenamefont {Young}}]{Fu20}%
  \BibitemOpen
  \bibfield  {author} {\bibinfo {author} {\bibfnamefont {S.-P.~P.}\
  \bibnamefont {Fu}}, \bibinfo {author} {\bibfnamefont {R.}~\bibnamefont
  {Ryham}}, \bibinfo {author} {\bibfnamefont {A.}~\bibnamefont {Kl\"ockner}},
  \bibinfo {author} {\bibfnamefont {M.}~\bibnamefont {Wala}}, \bibinfo {author}
  {\bibfnamefont {S.}~\bibnamefont {Jiang}},\ and\ \bibinfo {author}
  {\bibfnamefont {Y.-N.}\ \bibnamefont {Young}},\ }\href@noop {} {\bibfield
  {journal} {\bibinfo  {journal} {Multiscale Modeling \& Simulation}\ }\textbf
  {\bibinfo {volume} {18}},\ \bibinfo {pages} {79} (\bibinfo {year}
  {2020})}\BibitemShut {NoStop}%
\bibitem [{\citenamefont {Fu}\ \emph {et~al.}(2022)\citenamefont {Fu},
  \citenamefont {Quaife}, \citenamefont {Ryham},\ and\ \citenamefont
  {Young}}]{Fu2022_JFM}%
  \BibitemOpen
  \bibfield  {author} {\bibinfo {author} {\bibfnamefont {S.-P.}\ \bibnamefont
  {Fu}}, \bibinfo {author} {\bibfnamefont {B.}~\bibnamefont {Quaife}}, \bibinfo
  {author} {\bibfnamefont {R.}~\bibnamefont {Ryham}},\ and\ \bibinfo {author}
  {\bibfnamefont {Y.-N.}\ \bibnamefont {Young}},\ }\href@noop {} {\bibfield
  {journal} {\bibinfo  {journal} {Journal of Fluid Mechanics}\ }\textbf
  {\bibinfo {volume} {941}},\ \bibinfo {pages} {A41} (\bibinfo {year}
  {2022})}\BibitemShut {NoStop}%
\bibitem [{\citenamefont {Nagle}\ and\ \citenamefont
  {Tristram-Nagle}(2000)}]{NaTr00}%
  \BibitemOpen
  \bibfield  {author} {\bibinfo {author} {\bibfnamefont {J.~F.}\ \bibnamefont
  {Nagle}}\ and\ \bibinfo {author} {\bibfnamefont {S.}~\bibnamefont
  {Tristram-Nagle}},\ }\href@noop {} {\bibfield  {journal} {\bibinfo  {journal}
  {Biochimica et Biophysica Acta (BBA) - Reviews on Biomembranes}\ }\textbf
  {\bibinfo {volume} {1469}},\ \bibinfo {pages} {159} (\bibinfo {year}
  {2000})}\BibitemShut {NoStop}%
\bibitem [{\citenamefont {Kr\"uger}\ \emph {et~al.}(2013)\citenamefont
  {Kr\"uger}, \citenamefont {Frijters}, \citenamefont {G\"unther},
  \citenamefont {Kaoui},\ and\ \citenamefont {Harting}}]{KrFiGuKaHa13}%
  \BibitemOpen
  \bibfield  {author} {\bibinfo {author} {\bibfnamefont {T.}~\bibnamefont
  {Kr\"uger}}, \bibinfo {author} {\bibfnamefont {S.}~\bibnamefont {Frijters}},
  \bibinfo {author} {\bibfnamefont {F.}~\bibnamefont {G\"unther}}, \bibinfo
  {author} {\bibfnamefont {B.}~\bibnamefont {Kaoui}},\ and\ \bibinfo {author}
  {\bibfnamefont {J.}~\bibnamefont {Harting}},\ }\href@noop {} {\bibfield
  {journal} {\bibinfo  {journal} {The European Physical Journal Special
  Topics}\ }\textbf {\bibinfo {volume} {222}},\ \bibinfo {pages} {177}
  (\bibinfo {year} {2013})}\BibitemShut {NoStop}%
\bibitem [{\citenamefont {Grandmaison}\ \emph {et~al.}(2021)\citenamefont
  {Grandmaison}, \citenamefont {Brancherie},\ and\ \citenamefont
  {Salsac}}]{grandmaison_brancherie_salsac_2021}%
  \BibitemOpen
  \bibfield  {author} {\bibinfo {author} {\bibfnamefont {N.}~\bibnamefont
  {Grandmaison}}, \bibinfo {author} {\bibfnamefont {D.}~\bibnamefont
  {Brancherie}},\ and\ \bibinfo {author} {\bibfnamefont {A.-V.}\ \bibnamefont
  {Salsac}},\ }\href {https://doi.org/10.1017/jfm.2020.652} {\bibfield
  {journal} {\bibinfo  {journal} {Journal of Fluid Mechanics}\ }\textbf
  {\bibinfo {volume} {914}},\ \bibinfo {pages} {A25} (\bibinfo {year}
  {2021})}\BibitemShut {NoStop}%
\bibitem [{\citenamefont {Er-Rafik}\ \emph {et~al.}(2022)\citenamefont
  {Er-Rafik}, \citenamefont {Ferji}, \citenamefont {Combet}, \citenamefont
  {Sandre}, \citenamefont {Lecommandoux}, \citenamefont {Schmutz},
  \citenamefont {Le~Meins},\ and\ \citenamefont {Marques}}]{D2SM00179A}%
  \BibitemOpen
  \bibfield  {author} {\bibinfo {author} {\bibfnamefont {M.}~\bibnamefont
  {Er-Rafik}}, \bibinfo {author} {\bibfnamefont {K.}~\bibnamefont {Ferji}},
  \bibinfo {author} {\bibfnamefont {J.}~\bibnamefont {Combet}}, \bibinfo
  {author} {\bibfnamefont {O.}~\bibnamefont {Sandre}}, \bibinfo {author}
  {\bibfnamefont {S.}~\bibnamefont {Lecommandoux}}, \bibinfo {author}
  {\bibfnamefont {M.}~\bibnamefont {Schmutz}}, \bibinfo {author} {\bibfnamefont
  {J.-F.}\ \bibnamefont {Le~Meins}},\ and\ \bibinfo {author} {\bibfnamefont
  {C.~M.}\ \bibnamefont {Marques}},\ }\href@noop {} {\bibfield  {journal}
  {\bibinfo  {journal} {Soft Matter}\ }\textbf {\bibinfo {volume} {18}},\
  \bibinfo {pages} {3318} (\bibinfo {year} {2022})}\BibitemShut {NoStop}%
\bibitem [{\citenamefont {Keller}\ and\ \citenamefont
  {Skalak}(1982)}]{keller_skalak_1982}%
  \BibitemOpen
  \bibfield  {author} {\bibinfo {author} {\bibfnamefont {S.~R.}\ \bibnamefont
  {Keller}}\ and\ \bibinfo {author} {\bibfnamefont {R.}~\bibnamefont
  {Skalak}},\ }\href@noop {} {\bibfield  {journal} {\bibinfo  {journal}
  {Journal of Fluid Mechanics}\ }\textbf {\bibinfo {volume} {120}},\ \bibinfo
  {pages} {27} (\bibinfo {year} {1982})}\BibitemShut {NoStop}%
\bibitem [{\citenamefont {Finken}\ \emph {et~al.}(2008)\citenamefont {Finken},
  \citenamefont {Lamura}, \citenamefont {Seifert},\ and\ \citenamefont
  {Gompper}}]{Finken08}%
  \BibitemOpen
  \bibfield  {author} {\bibinfo {author} {\bibfnamefont {R.}~\bibnamefont
  {Finken}}, \bibinfo {author} {\bibfnamefont {A.}~\bibnamefont {Lamura}},
  \bibinfo {author} {\bibfnamefont {U.}~\bibnamefont {Seifert}},\ and\ \bibinfo
  {author} {\bibfnamefont {G.}~\bibnamefont {Gompper}},\ }\href@noop {}
  {\bibfield  {journal} {\bibinfo  {journal} {The European Physical Journal E}\
  }\textbf {\bibinfo {volume} {25}},\ \bibinfo {pages} {309} (\bibinfo {year}
  {2008})}\BibitemShut {NoStop}%
\bibitem [{\citenamefont {Zhao}\ and\ \citenamefont
  {Shaqfeh}(2011)}]{Shaqfeh11}%
  \BibitemOpen
  \bibfield  {author} {\bibinfo {author} {\bibfnamefont {H.}~\bibnamefont
  {Zhao}}\ and\ \bibinfo {author} {\bibfnamefont {E.~S.~G.}\ \bibnamefont
  {Shaqfeh}},\ }\href@noop {} {\bibfield  {journal} {\bibinfo  {journal}
  {Journal of Fluid Mechanics}\ }\textbf {\bibinfo {volume} {674}},\ \bibinfo
  {pages} {578} (\bibinfo {year} {2011})}\BibitemShut {NoStop}%
\bibitem [{\citenamefont {Brandner}\ \emph {et~al.}(2019)\citenamefont
  {Brandner}, \citenamefont {Timr}, \citenamefont {Melchionna}, \citenamefont
  {Derreumaux}, \citenamefont {Baaden},\ and\ \citenamefont
  {Sterpone}}]{Brandner2019}%
  \BibitemOpen
  \bibfield  {author} {\bibinfo {author} {\bibfnamefont {A.~F.}\ \bibnamefont
  {Brandner}}, \bibinfo {author} {\bibfnamefont {S.}~\bibnamefont {Timr}},
  \bibinfo {author} {\bibfnamefont {S.}~\bibnamefont {Melchionna}}, \bibinfo
  {author} {\bibfnamefont {P.}~\bibnamefont {Derreumaux}}, \bibinfo {author}
  {\bibfnamefont {M.}~\bibnamefont {Baaden}},\ and\ \bibinfo {author}
  {\bibfnamefont {F.}~\bibnamefont {Sterpone}},\ }\href@noop {} {\bibfield
  {journal} {\bibinfo  {journal} {Scientific Reports}\ }\textbf {\bibinfo
  {volume} {9}},\ \bibinfo {pages} {16450} (\bibinfo {year}
  {2019})}\BibitemShut {NoStop}%
\bibitem [{\citenamefont {Banik}\ \emph {et~al.}(2021)\citenamefont {Banik},
  \citenamefont {Sett}, \citenamefont {Bakli}, \citenamefont {Raychaudhuri},
  \citenamefont {Chakraborty},\ and\ \citenamefont
  {Mukherjee}}]{Baniketal2021}%
  \BibitemOpen
  \bibfield  {author} {\bibinfo {author} {\bibfnamefont {M.}~\bibnamefont
  {Banik}}, \bibinfo {author} {\bibfnamefont {S.}~\bibnamefont {Sett}},
  \bibinfo {author} {\bibfnamefont {C.}~\bibnamefont {Bakli}}, \bibinfo
  {author} {\bibfnamefont {A.~K.}\ \bibnamefont {Raychaudhuri}}, \bibinfo
  {author} {\bibfnamefont {S.}~\bibnamefont {Chakraborty}},\ and\ \bibinfo
  {author} {\bibfnamefont {R.}~\bibnamefont {Mukherjee}},\ }\href@noop {}
  {\bibfield  {journal} {\bibinfo  {journal} {Scientific Reports}\ }\textbf
  {\bibinfo {volume} {11}},\ \bibinfo {pages} {1} (\bibinfo {year}
  {2021})}\BibitemShut {NoStop}%
\bibitem [{\citenamefont {Hu}\ \emph {et~al.}(2019)\citenamefont {Hu},
  \citenamefont {Sun}, \citenamefont {Zhu}, \citenamefont {Huang},
  \citenamefont {Li},\ and\ \citenamefont {Sun}}]{C9NR05885K}%
  \BibitemOpen
  \bibfield  {author} {\bibinfo {author} {\bibfnamefont {F.-F.}\ \bibnamefont
  {Hu}}, \bibinfo {author} {\bibfnamefont {Y.-W.}\ \bibnamefont {Sun}},
  \bibinfo {author} {\bibfnamefont {Y.-L.}\ \bibnamefont {Zhu}}, \bibinfo
  {author} {\bibfnamefont {Y.-N.}\ \bibnamefont {Huang}}, \bibinfo {author}
  {\bibfnamefont {Z.-W.}\ \bibnamefont {Li}},\ and\ \bibinfo {author}
  {\bibfnamefont {Z.-Y.}\ \bibnamefont {Sun}},\ }\href@noop {} {\bibfield
  {journal} {\bibinfo  {journal} {Nanoscale}\ }\textbf {\bibinfo {volume}
  {11}},\ \bibinfo {pages} {17350} (\bibinfo {year} {2019})}\BibitemShut
  {NoStop}%
\bibitem [{\citenamefont {Mar\v{c}elja}(1977)}]{Ma77}%
  \BibitemOpen
  \bibfield  {author} {\bibinfo {author} {\bibfnamefont {S.}~\bibnamefont
  {Mar\v{c}elja}},\ }\href@noop {} {\bibfield  {journal} {\bibinfo  {journal}
  {Croatica Chemica Acta}\ }\textbf {\bibinfo {volume} {49}},\ \bibinfo {pages}
  {347} (\bibinfo {year} {1977})}\BibitemShut {NoStop}%
\bibitem [{\citenamefont {Gompper}\ \emph {et~al.}(1994)\citenamefont
  {Gompper}, \citenamefont {Hauser},\ and\ \citenamefont
  {Kornyshev}}]{GoHaKo94}%
  \BibitemOpen
  \bibfield  {author} {\bibinfo {author} {\bibfnamefont {G.}~\bibnamefont
  {Gompper}}, \bibinfo {author} {\bibfnamefont {M.}~\bibnamefont {Hauser}},\
  and\ \bibinfo {author} {\bibfnamefont {A.~A.}\ \bibnamefont {Kornyshev}},\
  }\href@noop {} {\bibfield  {journal} {\bibinfo  {journal} {The Journal of
  Chemical Physics}\ }\textbf {\bibinfo {volume} {101}},\ \bibinfo {pages}
  {3378} (\bibinfo {year} {1994})}\BibitemShut {NoStop}%
\bibitem [{\citenamefont {Eriksson}\ \emph {et~al.}(1989)\citenamefont
  {Eriksson}, \citenamefont {Ljunggren},\ and\ \citenamefont
  {Claesson}}]{ErLjCl89}%
  \BibitemOpen
  \bibfield  {author} {\bibinfo {author} {\bibfnamefont {J.~C.}\ \bibnamefont
  {Eriksson}}, \bibinfo {author} {\bibfnamefont {S.}~\bibnamefont
  {Ljunggren}},\ and\ \bibinfo {author} {\bibfnamefont {P.~M.}\ \bibnamefont
  {Claesson}},\ }\href@noop {} {\bibfield  {journal} {\bibinfo  {journal} {J.
  Chem. Soc.{,} Faraday Trans. 2}\ }\textbf {\bibinfo {volume} {85}},\ \bibinfo
  {pages} {163} (\bibinfo {year} {1989})}\BibitemShut {NoStop}%
\bibitem [{\citenamefont {Lin}\ \emph {et~al.}(2005)\citenamefont {Lin},
  \citenamefont {Meyer}, \citenamefont {Tadmor}, \citenamefont
  {Israelachvili},\ and\ \citenamefont {Kuhl}}]{Lietal05}%
  \BibitemOpen
  \bibfield  {author} {\bibinfo {author} {\bibfnamefont {Q.}~\bibnamefont
  {Lin}}, \bibinfo {author} {\bibfnamefont {E.~E.}\ \bibnamefont {Meyer}},
  \bibinfo {author} {\bibfnamefont {M.}~\bibnamefont {Tadmor}}, \bibinfo
  {author} {\bibfnamefont {J.~N.}\ \bibnamefont {Israelachvili}},\ and\
  \bibinfo {author} {\bibfnamefont {T.~L.}\ \bibnamefont {Kuhl}},\ }\href@noop
  {} {\bibfield  {journal} {\bibinfo  {journal} {Langmuir}\ }\textbf {\bibinfo
  {volume} {21}},\ \bibinfo {pages} {251} (\bibinfo {year} {2005})}\BibitemShut
  {NoStop}%
\bibitem [{\citenamefont {Israelachvili}\ \emph {et~al.}(1980)\citenamefont
  {Israelachvili}, \citenamefont {Mar\v{c}elja},\ and\ \citenamefont
  {Horn}}]{Israelachvili80}%
  \BibitemOpen
  \bibfield  {author} {\bibinfo {author} {\bibfnamefont {J.~N.}\ \bibnamefont
  {Israelachvili}}, \bibinfo {author} {\bibfnamefont {S.}~\bibnamefont
  {Mar\v{c}elja}},\ and\ \bibinfo {author} {\bibfnamefont {R.~G.}\ \bibnamefont
  {Horn}},\ }\href@noop {} {\bibfield  {journal} {\bibinfo  {journal}
  {Quarterly Reviews of Biophysics}\ }\textbf {\bibinfo {volume} {13}},\
  \bibinfo {pages} {121} (\bibinfo {year} {1980})}\BibitemShut {NoStop}%
\bibitem [{\citenamefont {Kohl}\ \emph {et~al.}(2022)\citenamefont {Kohl},
  \citenamefont {Corona}, \citenamefont {Cheruvu},\ and\ \citenamefont
  {Veerapaneni}}]{kohl-cor-che-vee22}%
  \BibitemOpen
  \bibfield  {author} {\bibinfo {author} {\bibfnamefont {R.}~\bibnamefont
  {Kohl}}, \bibinfo {author} {\bibfnamefont {E.}~\bibnamefont {Corona}},
  \bibinfo {author} {\bibfnamefont {V.}~\bibnamefont {Cheruvu}},\ and\ \bibinfo
  {author} {\bibfnamefont {S.}~\bibnamefont {Veerapaneni}},\ }\href@noop {}
  {\bibfield  {journal} {\bibinfo  {journal} {arxiv}\ }\textbf {\bibinfo
  {volume} {2104.14068}} (\bibinfo {year} {2022})}\BibitemShut {NoStop}%
\bibitem [{\citenamefont {Hsiao}\ and\ \citenamefont
  {Wendland}(2008)}]{Hsiao2008}%
  \BibitemOpen
  \bibfield  {author} {\bibinfo {author} {\bibfnamefont {G.~C.}\ \bibnamefont
  {Hsiao}}\ and\ \bibinfo {author} {\bibfnamefont {W.~L.}\ \bibnamefont
  {Wendland}},\ }\href@noop {} {\emph {\bibinfo {title} {{Boundary Integral
  Equations}}}}\ (\bibinfo  {publisher} {Springer Berlin Heidelberg},\ \bibinfo
  {address} {Berlin, Heidelberg},\ \bibinfo {year} {2008})\BibitemShut
  {NoStop}%
\bibitem [{\citenamefont {Power}\ and\ \citenamefont
  {Miranda}(1987)}]{pow-mir1987}%
  \BibitemOpen
  \bibfield  {author} {\bibinfo {author} {\bibfnamefont {H.}~\bibnamefont
  {Power}}\ and\ \bibinfo {author} {\bibfnamefont {G.}~\bibnamefont
  {Miranda}},\ }\href@noop {} {\bibfield  {journal} {\bibinfo  {journal} {SIAM
  Journal on Applied Mathematics}\ }\textbf {\bibinfo {volume} {47}},\ \bibinfo
  {pages} {689} (\bibinfo {year} {1987})}\BibitemShut {NoStop}%
\bibitem [{\citenamefont {Pozrikidis}(1992)}]{poz1992}%
  \BibitemOpen
  \bibfield  {author} {\bibinfo {author} {\bibfnamefont {C.}~\bibnamefont
  {Pozrikidis}},\ }\href@noop {} {\emph {\bibinfo {title} {Boundary Integral
  and Singularity Methods for Linearized Viscous Flow}}}\ (\bibinfo
  {publisher} {Cambridge University Press},\ \bibinfo {address} {New York, NY,
  USA},\ \bibinfo {year} {1992})\BibitemShut {NoStop}%
\bibitem [{\citenamefont {Rachh}\ and\ \citenamefont
  {Greengard}(2016)}]{rac-gre2016}%
  \BibitemOpen
  \bibfield  {author} {\bibinfo {author} {\bibfnamefont {M.}~\bibnamefont
  {Rachh}}\ and\ \bibinfo {author} {\bibfnamefont {L.}~\bibnamefont
  {Greengard}},\ }\href@noop {} {\bibfield  {journal} {\bibinfo  {journal}
  {SIAM Journal on Numerical Analysis}\ }\textbf {\bibinfo {volume} {54}},\
  \bibinfo {pages} {2889} (\bibinfo {year} {2016})}\BibitemShut {NoStop}%
\bibitem [{\citenamefont {Corona}\ \emph {et~al.}(2017)\citenamefont {Corona},
  \citenamefont {Greengard}, \citenamefont {Rachh},\ and\ \citenamefont
  {Veerapaneni}}]{cor-gre-rac-vee2017}%
  \BibitemOpen
  \bibfield  {author} {\bibinfo {author} {\bibfnamefont {E.}~\bibnamefont
  {Corona}}, \bibinfo {author} {\bibfnamefont {L.}~\bibnamefont {Greengard}},
  \bibinfo {author} {\bibfnamefont {M.}~\bibnamefont {Rachh}},\ and\ \bibinfo
  {author} {\bibfnamefont {S.}~\bibnamefont {Veerapaneni}},\ }\href@noop {}
  {\bibfield  {journal} {\bibinfo  {journal} {Journal of Computational
  Physics}\ }\textbf {\bibinfo {volume} {332}},\ \bibinfo {pages} {504}
  (\bibinfo {year} {2017})}\BibitemShut {NoStop}%
\bibitem [{\citenamefont {Quaife}\ and\ \citenamefont
  {Biros}(2014)}]{qua-bir2014}%
  \BibitemOpen
  \bibfield  {author} {\bibinfo {author} {\bibfnamefont {B.}~\bibnamefont
  {Quaife}}\ and\ \bibinfo {author} {\bibfnamefont {G.}~\bibnamefont {Biros}},\
  }\href@noop {} {\bibfield  {journal} {\bibinfo  {journal} {Journal of
  Computational Physics}\ }\textbf {\bibinfo {volume} {274}},\ \bibinfo {pages}
  {245} (\bibinfo {year} {2014})}\BibitemShut {NoStop}%
\bibitem [{\citenamefont {Boal}(2012)}]{Boal}%
  \BibitemOpen
  \bibfield  {author} {\bibinfo {author} {\bibfnamefont {D.}~\bibnamefont
  {Boal}},\ }\href@noop {} {\emph {\bibinfo {title} {Mechanics of the Cell
  Second Edition}}}\ (\bibinfo  {publisher} {Cambridge University Press},\
  \bibinfo {address} {Cambridge, United Kingdom},\ \bibinfo {year}
  {2012})\BibitemShut {NoStop}%
\bibitem [{\citenamefont {Kuzmin}\ \emph {et~al.}(2005)\citenamefont {Kuzmin},
  \citenamefont {Akimov}, \citenamefont {Chizmadzhev}, \citenamefont
  {Zimmerberg},\ and\ \citenamefont {Cohen}}]{KUZMIN2005}%
  \BibitemOpen
  \bibfield  {author} {\bibinfo {author} {\bibfnamefont {P.~I.}\ \bibnamefont
  {Kuzmin}}, \bibinfo {author} {\bibfnamefont {S.~A.}\ \bibnamefont {Akimov}},
  \bibinfo {author} {\bibfnamefont {Y.~A.}\ \bibnamefont {Chizmadzhev}},
  \bibinfo {author} {\bibfnamefont {J.}~\bibnamefont {Zimmerberg}},\ and\
  \bibinfo {author} {\bibfnamefont {F.~S.}\ \bibnamefont {Cohen}},\ }\href@noop
  {} {\bibfield  {journal} {\bibinfo  {journal} {Biophysical Journal}\ }\textbf
  {\bibinfo {volume} {88}},\ \bibinfo {pages} {1120} (\bibinfo {year}
  {2005})}\BibitemShut {NoStop}%
\bibitem [{\citenamefont {Petelska}(2012)}]{Petelska2012}%
  \BibitemOpen
  \bibfield  {author} {\bibinfo {author} {\bibfnamefont {A.~D.}\ \bibnamefont
  {Petelska}},\ }\href@noop {} {\bibfield  {journal} {\bibinfo  {journal}
  {Central European Journal of Chemistry}\ }\textbf {\bibinfo {volume} {10}},\
  \bibinfo {pages} {16} (\bibinfo {year} {2012})}\BibitemShut {NoStop}%
\bibitem [{\citenamefont {Jackson}(2016)}]{Jackson2016}%
  \BibitemOpen
  \bibfield  {author} {\bibinfo {author} {\bibfnamefont {M.~B.}\ \bibnamefont
  {Jackson}},\ }\href@noop {} {\bibfield  {journal} {\bibinfo  {journal}
  {Scientific Reports}\ }\textbf {\bibinfo {volume} {6}},\ \bibinfo {pages} {6}
  (\bibinfo {year} {2016})}\BibitemShut {NoStop}%
\bibitem [{\citenamefont {Garc\'{i}a-S\'{a}ez}\ \emph
  {et~al.}(2007)\citenamefont {Garc\'{i}a-S\'{a}ez}, \citenamefont {Chiantia},\
  and\ \citenamefont {Schwille}}]{GarciaSaez}%
  \BibitemOpen
  \bibfield  {author} {\bibinfo {author} {\bibfnamefont {A.~J.}\ \bibnamefont
  {Garc\'{i}a-S\'{a}ez}}, \bibinfo {author} {\bibfnamefont {S.}~\bibnamefont
  {Chiantia}},\ and\ \bibinfo {author} {\bibfnamefont {P.}~\bibnamefont
  {Schwille}},\ }\href@noop {} {\bibfield  {journal} {\bibinfo  {journal}
  {Journal of Biological Chemistry}\ }\textbf {\bibinfo {volume} {282}},\
  \bibinfo {pages} {33537} (\bibinfo {year} {2007})}\BibitemShut {NoStop}%
\bibitem [{\citenamefont {Mar\v{c}elja}\ and\ \citenamefont
  {Radic}(1976)}]{MaRa76}%
  \BibitemOpen
  \bibfield  {author} {\bibinfo {author} {\bibfnamefont {S.}~\bibnamefont
  {Mar\v{c}elja}}\ and\ \bibinfo {author} {\bibfnamefont {N.}~\bibnamefont
  {Radic}},\ }\href@noop {} {\bibfield  {journal} {\bibinfo  {journal}
  {Chemical Physical Letters}\ }\textbf {\bibinfo {volume} {42}},\ \bibinfo
  {pages} {129} (\bibinfo {year} {1976})}\BibitemShut {NoStop}%
\bibitem [{\citenamefont {Selinger}(2016)}]{Selinger2016}%
  \BibitemOpen
  \bibfield  {author} {\bibinfo {author} {\bibfnamefont {J.~V.}\ \bibnamefont
  {Selinger}},\ }\href@noop {} {\emph {\bibinfo {title} {Introduction to the
  Theory of Soft Matter: From Ideal Gases to Liquid Crystals}}}\ (\bibinfo
  {publisher} {Springer International Publishing},\ \bibinfo {address}
  {Switzerland},\ \bibinfo {year} {2016})\BibitemShut {NoStop}%
\bibitem [{\citenamefont {Du}\ \emph {et~al.}(2016)\citenamefont {Du},
  \citenamefont {Fu}, \citenamefont {Zhu}, \citenamefont {Ma},\ and\
  \citenamefont {Li}}]{DuFuZhuMaLi2016_AICHEJ}%
  \BibitemOpen
  \bibfield  {author} {\bibinfo {author} {\bibfnamefont {W.}~\bibnamefont
  {Du}}, \bibinfo {author} {\bibfnamefont {T.}~\bibnamefont {Fu}}, \bibinfo
  {author} {\bibfnamefont {C.}~\bibnamefont {Zhu}}, \bibinfo {author}
  {\bibfnamefont {Y.}~\bibnamefont {Ma}},\ and\ \bibinfo {author}
  {\bibfnamefont {H.~Z.}\ \bibnamefont {Li}},\ }\href@noop {} {\bibfield
  {journal} {\bibinfo  {journal} {AIChE J.}\ }\textbf {\bibinfo {volume}
  {62}},\ \bibinfo {pages} {325} (\bibinfo {year} {2016})}\BibitemShut
  {NoStop}%
\bibitem [{\citenamefont {Stone}(1994)}]{Stone1994_ARFM}%
  \BibitemOpen
  \bibfield  {author} {\bibinfo {author} {\bibfnamefont {H.~A.}\ \bibnamefont
  {Stone}},\ }\href@noop {} {\bibfield  {journal} {\bibinfo  {journal} {Annu.
  Rev. Fluid Mech.}\ }\textbf {\bibinfo {volume} {26}},\ \bibinfo {pages} {65}
  (\bibinfo {year} {1994})}\BibitemShut {NoStop}%
\bibitem [{\citenamefont {Schwalbe}\ \emph {et~al.}(2010)\citenamefont
  {Schwalbe}, \citenamefont {Vlahovska},\ and\ \citenamefont
  {Miksis}}]{sch-vla-mik2010}%
  \BibitemOpen
  \bibfield  {author} {\bibinfo {author} {\bibfnamefont {J.~T.}\ \bibnamefont
  {Schwalbe}}, \bibinfo {author} {\bibfnamefont {P.~M.}\ \bibnamefont
  {Vlahovska}},\ and\ \bibinfo {author} {\bibfnamefont {M.~J.}\ \bibnamefont
  {Miksis}},\ }\href@noop {} {\bibfield  {journal} {\bibinfo  {journal}
  {Journal of Fluid Mechanics}\ }\textbf {\bibinfo {volume} {647}},\ \bibinfo
  {pages} {403} (\bibinfo {year} {2010})}\BibitemShut {NoStop}%
\bibitem [{\citenamefont {den Otter}\ and\ \citenamefont
  {Shkulipa}(2007)}]{denOtter2007}%
  \BibitemOpen
  \bibfield  {author} {\bibinfo {author} {\bibfnamefont {W.~K.}\ \bibnamefont
  {den Otter}}\ and\ \bibinfo {author} {\bibfnamefont {S.~A.}\ \bibnamefont
  {Shkulipa}},\ }\href@noop {} {\bibfield  {journal} {\bibinfo  {journal}
  {Biophysical Journal}\ }\textbf {\bibinfo {volume} {93}},\ \bibinfo {pages}
  {423} (\bibinfo {year} {2007})}\BibitemShut {NoStop}%
\bibitem [{\citenamefont {Zgorski}\ \emph {et~al.}(2019)\citenamefont
  {Zgorski}, \citenamefont {Pastor},\ and\ \citenamefont
  {Lyman}}]{Zgorski2019}%
  \BibitemOpen
  \bibfield  {author} {\bibinfo {author} {\bibfnamefont {A.}~\bibnamefont
  {Zgorski}}, \bibinfo {author} {\bibfnamefont {R.~W.}\ \bibnamefont
  {Pastor}},\ and\ \bibinfo {author} {\bibfnamefont {E.}~\bibnamefont
  {Lyman}},\ }\href@noop {} {\bibfield  {journal} {\bibinfo  {journal} {Journal
  of Chemical Theory and Computation}\ }\textbf {\bibinfo {volume} {15}},\
  \bibinfo {pages} {6471} (\bibinfo {year} {2019})}\BibitemShut {NoStop}%
\bibitem [{\citenamefont {Chabanon}\ \emph {et~al.}(2017)\citenamefont
  {Chabanon}, \citenamefont {Ho}, \citenamefont {Liedberg}, \citenamefont
  {Parikh},\ and\ \citenamefont {Rangamani}}]{chabanon2017}%
  \BibitemOpen
  \bibfield  {author} {\bibinfo {author} {\bibfnamefont {M.}~\bibnamefont
  {Chabanon}}, \bibinfo {author} {\bibfnamefont {J.~C.~S.}\ \bibnamefont {Ho}},
  \bibinfo {author} {\bibfnamefont {B.}~\bibnamefont {Liedberg}}, \bibinfo
  {author} {\bibfnamefont {A.~N.}\ \bibnamefont {Parikh}},\ and\ \bibinfo
  {author} {\bibfnamefont {P.}~\bibnamefont {Rangamani}},\ }\href@noop {}
  {\bibfield  {journal} {\bibinfo  {journal} {Biophysical Journal}\ }\textbf
  {\bibinfo {volume} {112}},\ \bibinfo {pages} {1682} (\bibinfo {year}
  {2017})}\BibitemShut {NoStop}%
\bibitem [{\citenamefont {Quaife}\ \emph {et~al.}(2021)\citenamefont {Quaife},
  \citenamefont {Gannon},\ and\ \citenamefont {Young}}]{qua-gan-you2021}%
  \BibitemOpen
  \bibfield  {author} {\bibinfo {author} {\bibfnamefont {B.}~\bibnamefont
  {Quaife}}, \bibinfo {author} {\bibfnamefont {A.}~\bibnamefont {Gannon}},\
  and\ \bibinfo {author} {\bibfnamefont {Y.-N.}\ \bibnamefont {Young}},\
  }\href@noop {} {\bibfield  {journal} {\bibinfo  {journal} {Physical Review
  Fluids}\ }\textbf {\bibinfo {volume} {6}},\ \bibinfo {pages} {073601}
  (\bibinfo {year} {2021})}\BibitemShut {NoStop}%
\bibitem [{\citenamefont {Konijn}\ \emph {et~al.}(2014)\citenamefont {Konijn},
  \citenamefont {Sanderink},\ and\ \citenamefont {Kruyt}}]{KONIJN201461}%
  \BibitemOpen
  \bibfield  {author} {\bibinfo {author} {\bibfnamefont {B.}~\bibnamefont
  {Konijn}}, \bibinfo {author} {\bibfnamefont {O.}~\bibnamefont {Sanderink}},\
  and\ \bibinfo {author} {\bibfnamefont {N.}~\bibnamefont {Kruyt}},\
  }\href@noop {} {\bibfield  {journal} {\bibinfo  {journal} {Powder
  Technology}\ }\textbf {\bibinfo {volume} {266}},\ \bibinfo {pages} {61}
  (\bibinfo {year} {2014})}\BibitemShut {NoStop}%
\bibitem [{\citenamefont {Amador}\ \emph {et~al.}(2021)\citenamefont {Amador},
  \citenamefont {van Dijk}, \citenamefont {Kieffer}, \citenamefont
  {Aubin-Tam},\ and\ \citenamefont {Tam}}]{doi:10.1073/pnas.2100156118}%
  \BibitemOpen
  \bibfield  {author} {\bibinfo {author} {\bibfnamefont {G.~J.}\ \bibnamefont
  {Amador}}, \bibinfo {author} {\bibfnamefont {D.}~\bibnamefont {van Dijk}},
  \bibinfo {author} {\bibfnamefont {R.}~\bibnamefont {Kieffer}}, \bibinfo
  {author} {\bibfnamefont {M.-E.}\ \bibnamefont {Aubin-Tam}},\ and\ \bibinfo
  {author} {\bibfnamefont {D.}~\bibnamefont {Tam}},\ }\href@noop {} {\bibfield
  {journal} {\bibinfo  {journal} {Proceedings of the National Academy of
  Sciences}\ }\textbf {\bibinfo {volume} {118}},\ \bibinfo {pages}
  {e2100156118} (\bibinfo {year} {2021})}\BibitemShut {NoStop}%
\bibitem [{\citenamefont {McGlasson}\ and\ \citenamefont
  {Bradley}(2021)}]{McGlassonBradley2021}%
  \BibitemOpen
  \bibfield  {author} {\bibinfo {author} {\bibfnamefont {A.}~\bibnamefont
  {McGlasson}}\ and\ \bibinfo {author} {\bibfnamefont {L.~C.}\ \bibnamefont
  {Bradley}},\ }\href@noop {} {\bibfield  {journal} {\bibinfo  {journal}
  {Small}\ }\textbf {\bibinfo {volume} {17}},\ \bibinfo {pages} {2104926}
  (\bibinfo {year} {2021})}\BibitemShut {NoStop}%
\bibitem [{\citenamefont {Mallory}\ \emph {et~al.}(2017)\citenamefont
  {Mallory}, \citenamefont {Alarcon}, \citenamefont {Cacciuto},\ and\
  \citenamefont {Valeriani}}]{Mallory2017}%
  \BibitemOpen
  \bibfield  {author} {\bibinfo {author} {\bibfnamefont {S.~A.}\ \bibnamefont
  {Mallory}}, \bibinfo {author} {\bibfnamefont {F.}~\bibnamefont {Alarcon}},
  \bibinfo {author} {\bibfnamefont {A.}~\bibnamefont {Cacciuto}},\ and\
  \bibinfo {author} {\bibfnamefont {C.}~\bibnamefont {Valeriani}},\ }\href@noop
  {} {\bibfield  {journal} {\bibinfo  {journal} {New Journal of Physics}\
  }\textbf {\bibinfo {volume} {19}},\ \bibinfo {pages} {125014} (\bibinfo
  {year} {2017})}\BibitemShut {NoStop}%
\end{thebibliography}

\thispagestyle{empty}

\newpage
{\Large \bf

  \noindent Supplementary Material\\

  \noindent 
 Effects of Tunable Hydrophobicity on the Collective Hydrodynamics of Janus Particles under Flows}\\

\noindent 
Szu-Pei Fu$^{1,*},$ 
Rolf Ryham$^{2},$ 
Bryan Quaife$^{3}$ and Y.-N. Young$^{4},$
\\

\noindent
$^{1}$Department of Mathematics, Trinity College, Hartford, Connecticut 06106, USA

\noindent
$^{2}$Department of Mathematics, Fordham University, Bronx, NY, USA

\noindent
$^{3}$Department of Scientific Computing, Florida State University, Tallahassee, Florida 32306, USA

\noindent
$^{4}$Department of Mathematical Sciences, New Jersey Institute of Technology, Newark, NJ 07102 USA
\\

\noindent $^*$Corresponding author. Address: Department of Mathematics, Trinity College, 
300 Summit Street, Hartford, CT 06106. email: \text{peter.fu@trincoll.edu}

\setcounter{page}{1}

\setcounter{figure}{0}
\renewcommand{\thefigure}{S\arabic{figure}}

\setcounter{equation}{0}
\renewcommand{\theequation}{S\arabic{equation}}

\setcounter{section}{0}
\renewcommand{\thesection}{S\arabic{section}}



\sloppy
\section{Movie Captions}\mbox{} \\

\noindent
{\bf Movie S1. Relaxation} 
There are 60 circular particles with radius 1.25~nm that are initially
confined in a square box. The simulation results show the relaxation
with each of the three boundary conditions. The time step of all
simulations is $\Delta t=0.2$. The color on the boundary from blue to
red is for $\min g(\xx)$ to $\max g(\xx)$. All final configurations are
adopted in simulations with hydrodynamic flows. \\

\noindent
{\bf Movie S2. Structures in a Shear Flow without Ruptures} 
We adopt the relaxed configurations and place the JP structures in the
shear flow. For the choices of the shear rate, we pick: $\dot\gamma =
0.05$ for the vesicle, $\dot\gamma = 0.05$ for the bilayer, $\dot\gamma
= 0.05$ for the multi-lamellar, and $\dot\gamma = 0.1$ for the striated
configurations. The vesicle case undergoes tank-treading whereas the
initially disordered BC (i) case increases orientational order. The
multilamellar assemply behaves as a rigid body, and the striated
configuration moreso. No ruptures occur. \\

\noindent
{\bf Movie S3. Structures in a Taylor-Green Flow without Ruptures} 
We adopt the relaxed configurations and place the JP structures in the
Taylor-Green flow at low flow rates. For the choices of the flow
strength, we pick: $V_0 = 0.1$ for the vesicle, $V_0 = 0.1$ for the
disordered bilayer, $V_0 = 0.1$ for the multi-lamellar, and $V_0 = 0.2$
for the striated configurations. The vesicle stays inact, whereas the
dissordered bilayer is pulled apart. Like in Supplementary Movie S2, the
striated assembly is basically rigid. \\

\noindent
{\bf Movie S4. Structures in a Shear Flow with Ruptures} 
We adopt the relaxed configurations and place the JP structures in the
shear flow at a higher shear rate. For the choices of the shear rate, we
pick: $\dot\gamma = 0.075$ for the vesicle, $\dot\gamma = 0.1$ for the
bilayer, $\dot\gamma = 0.15$ for the multi-lamellar, and $\dot\gamma =
0.15$ for the striated configurations. The time step of all simulations
is $\Delta t=0.2$. In this movie, some clear structural ruptures occur
in each case. In order to observe the structure behaviors at later time,
we stabilize the frame by tracking the center of mass position of all
JP. \\

\noindent
{\bf Movie S5. Structures in a Taylor-Green Flow with Ruptures} 
We adopt the relaxed configurations and place the JP structures in the
Taylor-Green flow at higher flow rates. For the choices of the flow
strength, we pick: $V_0 = 0.2$ for the vesicle, $V_0 = 0.2$ for the
bilayer, $V_0 = 0.2$ for the multi-lamellar, and $V_0 = 0.3$ for the
striated configurations. Clear structural ruptures occur in each case.
In all cases, there is significant reduction in the orientational order
of the assemblies. While the BC (i) cases are broken into several
pieces, the main body of the BC (ii) and BC (iii) assemblies are not
pulled apart by the background flow.

\end{document}